\documentclass{aa}  
\usepackage{natbib}
\usepackage{comment}
\usepackage{graphicx}
\usepackage{txfonts}
\usepackage[dvipsnames]{xcolor}
\usepackage[rightcaption]{sidecap}
\newcommand{\A}{$A_{2}$}
\newcommand{\tb}{$t_{bar}$}
\newcommand{\Msun}{$\mbox{M}_{\sun}$}

\begin{document}
   \title{Simulating nearby disc galaxies on the main star formation sequence}
   \subtitle{I. Bar formation and the building of the central gas reservoir}

   \author{
   Pierrick Verwilghen \inst{\ref{origins},\ref{eso}} \and
   Eric~Emsellem\inst{\ref{eso},\ref{lyon}} \and 
   Florent Renaud\inst{\ref{sxb},\ref{usias}} \and
   Milena Valentini\inst{\ref{UniTs},\ref{OATs},\ref{ICSC}} \and
   Jiayi Sun\inst{\inst{\ref{Princeton}}} \and
   Sarah Jeffreson\inst{\ref{CFA}} \and
   Ralf S.\ Klessen\inst{\ref{ITA},\ref{IWR}} \and 
   Mattia~C.~Sormani\inst{\ref{insubria},\ref{DPUS}} \and
   Ashley.~T.~Barnes\inst{\ref{eso}} \and
   Klaus Dolag\inst{\ref{USM},\ref{MPAG}} \and
   Kathryn Grasha\inst{\ref{ANU}, \ref{ASTRO3D}} \and
   Fu-Heng Liang\inst{\ref{eso},\ref{Ox}} \and 
   Sharon Meidt\inst{\ref{SOU}} \and
   Justus Neumann\inst{\ref{MPIA}} \and
   Miguel Querejeta\inst{\ref{IGN}} \and
   Eva Schinnerer\inst{\ref{MPIA}} \and
   Thomas G. Williams\inst{\ref{Ox}}
   %
   }
   
\institute{
        Excellence Cluster ORIGINS, Boltzmannstraße 2,
        85748 Garching, Germany\label{origins}
        \and European Southern Observatory, Karl-Schwarzschild-Stra{\ss}e 2, 85748 Garching, Germany\label{eso}
        \and Univ Lyon, Univ Lyon1, ENS de Lyon, CNRS, Centre de Recherche Astrophysique de Lyon UMR5574, F-69230 Saint-Genis-Laval France\label{lyon}
        \and Observatoire Astronomique de Strasbourg, Universit\'e de Strasbourg, CNRS UMR 7550, F-67000 Strasbourg, France\label{sxb}
        \and University of Strasbourg Institute for Advanced Study, 5 all\'ee du G\'en\'eral Rouvillois, F-67083 Strasbourg, France \label{usias}
        \and Astronomy Unit, Department of Physics, University of Trieste, via Tiepolo 11, I-34131 Trieste, Italy\label{UniTs}
        \and INAF – Osservatorio Astronomico di Trieste, via Tiepolo 11, I-34131 Trieste, Italy\label{OATs}
        \and ICSC - Italian Research Center on High Performance Computing, Big Data and Quantum Computing\label{ICSC}
        \and Department of Astrophysical Sciences, Princeton University, 4 Ivy Lane, Princeton, NJ 08544, USA\label{Princeton}
        \and Center for Astrophysics, Harvard \& Smithsonian, 60 Garden Street, Cambridge MA, USA\label{CFA}
        \and Universit\"{a}t Heidelberg, Zentrum f\"{u}r Astronomie, Institut f\"{u}r Theoretische Astrophysik, Albert-Ueberle-Str. 2,\\ D-69120 Heidelberg, Germany \label{ITA} \and Universit\"{a}t Heidelberg, Interdisziplin\"{a}res Zentrum f\"{u}r Wissenschaftliches Rechnen, Im Neuenheimer Feld 205,\\ D-69120 Heidelberg, Germany\label{IWR}
        \and Universit{\`a} dell’Insubria, via Valleggio 11, 22100 Como, Italy\label{insubria}
        \and Department of Physics, University of Surrey, Guildford GU2 7XH, UK\label{DPUS}
        \and Universit\"{a}ts-Sternwarte, Fakult\"{a}t für Physik, Ludwig-Maximilians-Universit\"{a}t M\"{u}nchen, Scheinerstr.1, 81679 M\"{u}nchen, Germany\label{USM}
        \and Max-Planck-Institut f\"{u}r Astrophysik, Karl-Schwarzschild-Straße 1, 85741 Garching, Germany\label{MPAG}
        \and Research School of Astronomy and Astrophysics, Australian National University, Canberra, ACT 2611, Australia\label{ANU}
        \and ARC Centre of Excellence for All Sky Astrophysics in 3 Dimensions (ASTRO 3D), Australia\label{ASTRO3D}
        \and Sub-department of Astrophysics, Department of Physics, University of Oxford, Keble Road, Oxford OX1 3RH, UK\label{Ox}
        \and Sterrenkundig Observatorium, Universiteit Gent, Krijgslaan 281 S9, B-9000 Gent, Belgium\label{SOU}
        \and Max Planck Institut f\"ur Astronomie, K\"onigstuhl 17, 69117 Heidelberg, Germany\label{MPIA}
        \and Observatorio Astronómico Nacional (IGN), C/Alfonso XII 3, Madrid E-28014, Spain\label{IGN}
        }

   \date{Received 29 November, 2023; accepted 10 April, 2024}

 
  \abstract
   {Past studies have long emphasised the key role played by galactic stellar bars in the context of disc secular evolution, via the redistribution of gas and stars, the triggering of star formation, and the formation of prominent structures such as rings and central mass concentrations. However, the exact physical processes acting on those structures, as well as the timescales associated with the building and consumption of central gas reservoirs are still not well understood. We are building a suite of hydro-dynamical RAMSES simulations of isolated, low-redshift galaxies that mimic the properties of the PHANGS sample. The initial conditions of the models reproduce the observed stellar mass, disc scale length, or gas fraction, and this paper presents a first subset of these models. Most of our simulated galaxies develop a prominent bar structure, which itself triggers central gas fuelling and the building of an over-density with a typical scale of 100-1000 pc. We confirm that if the host galaxy features an ellipsoidal component, the formation of the bar and gas fuelling are delayed. We show that most of our simulations follow a common time evolution, when accounting for mass scaling and the bar formation time. In our simulations, the stellar mass of $10^{10}$~M$_{\odot}$ seems to mark a change in the phases describing the time evolution of the bar and its impact on the interstellar medium. In massive discs (M$_{\star} \geq 10^{10}$~M$_{\odot}$), we observe the formation of a central gas reservoir with star formation mostly occurring within a restricted starburst region, leading to a gas depletion phase. Lower-mass systems (M$_{\star} < 10^{10}$~M$_{\odot}$) do not exhibit such a depletion phase, and show a more homogeneous spread of star-forming regions along the bar structure, and do not appear to host inner bar-driven discs or rings. Our results seem to be supported by observations, and we briefly discuss how this new suite of simulations can help our understanding of the secular evolution of main sequence disc galaxies.}

   \keywords{Gas transport -- Gas reservoir --- Galactic dynamics -- bars --- Black holes}

   \maketitle
%
%
\nolinenumbers
\section{Introduction} \label{sec:intro}
Bars and spiral arms both form naturally in galaxy disc-like systems \citep[see][and references therein]{Eskridge2000, Buta2015, Sellwood2022}. They have a significant impact on the internal disc structures, induce specific torques and resonant regions, and impose, via the tumbling and varying gravitational potential, constraints on the galactic orbital skeleton. Bars have been specifically and extensively studied and modelled \citep{Combes1981, Kaufmann1996,Atha1992(bar), Kormendy2004, Barazza2008, Gadotti2008, Kraljic2012, Goz2015, Fragkoudi2017} and have a direct influence on the short- and long-term evolution of the system. We know that bars funnel the gas to the central regions ($\sim100$~pc) of galaxies, and have been often suggested to play a role in the large-scale fuelling of the central supermassive black hole \citep[SMBH;][]{Fukuda1998, Schlosman1989, Ho1997}.

Similarly to our own Milky Way, most nearby disc galaxies are also believed to host central SMBHs, with masses between 10$^{6}$ - 10$^{9}$ M$_{\odot}$, at their centres \citep{Magorrian1998, Reines2016}. When gas builds into a surrounding accretion disc, those SMBHs can drive nuclear activity \citep{Krolik1999, Padovani_2017}, releasing a large amount of energy in its environment in the form of mainly radiative (quasar mode) or mechanical (radio mode) feedback \citep{Sanders1988, Fabian2012}. Such active galactic nuclei (AGNs) may impact the dynamical evolution of their host galaxy, their chemical composition, and even modulate the central gas accretion itself \citep[see e.g.][and references therein]{Combes2017}.

The exact physical mechanisms and timelines associated with the fuelling of central SMBHs are still heavily debated \citep[see e.g. ][]{Hopkins2010, Cheung2015}, both due to the range of spatial and timescales involved and to the complexity and non-linear nature of the physical processes involved. The duty cycle of an accretion disc more directly depends on small-scale physics and the close environment of the black hole ($10^{-6}$ to $10^{-2}$~pc), while the larger-scale kiloparsec dynamics and structures may influence its long-term evolution. Two main avenues are generally called upon for the global transport of gas in disc galaxies. An external channel or origin is when low angular momentum gas is falling from, for example, the circum-galactic medium (CGM) towards the galaxy centre \citep{Lacey1985, Bilitewski2012}. An internal channel is connected to the secular evolution and processes, such as gravitational torques and stellar-driven feedback, acting as a way to extract angular momentum from the gas \citep{Porciani2002, Stewart2013, Uebler2014, Genel2015, Pezzulli2016, Schmidt2016, Valentini2017}. The required amount of gas to trigger the AGN is relatively small compared with the total mass of gas available in the galaxy \citep{Hickox2012, Padovani2017}. This means that only a very small fraction of the gas funnelled within the central few hundreds of parsecs is needed to account for the nuclear black-hole-related activity. Further considering the several orders of magnitude in spatial scale that are spanned from the galactic bar (kiloparsecs) to an accreting black hole \citep[$\sim$ $10^{-4} - 0.1$ pc, see e.g.][and references therein]{Guo23, Izumi23}, it is not yet clear how the bar relates to the processes that are most relevant to drive the duty cycle of AGNs.

Progress may come from the realisation that many disc galaxies host a central over-density of gas at a scale of a few hundreds of parsec from the centre \citep{Comeron2010}. While such structures may not directly relate with the ongoing nuclear activity itself, they can play the role of a `gas reservoir' and represent an intermediate milestone that we need to understand if we wish to draw a full picture of central gas accretion and its time evolution. Those gas reservoirs are often interpreted in the context of bars as coinciding with a change in the sign of the torques \citep{Sanders1977, Wada1994}, that is, the location of the so-called inner Lindblad resonance (ILR). However, these models fail to reproduce the results of hydro-dynamical simulations \citep{Sormani2015}, thus calling for a new framework that can explain the full picture \citep[see][]{Sormani2023}). In the Milky Way, a $\sim200$~pc central molecular zone (CMZ) with a gas mass of $\sim1 - 7 \times 10^7$~M$_{\odot}$ has been identified and extensively studied \citep{Ferriere2007, Longmore2013, Henshaw2023}. In other galaxies, such gas reservoirs (or CMZs) have also been observed and again, often interpreted in the context of the gas fuelling by bars, while details of its building and life cycle (consumption via e.g. star formation) are not well constrained.

Some pioneering work was carried out by \cite{Atha1992(gas)}, who studied the response of gas to an idealised two-dimensional barred potential and observed the building of a central gas concentration within the central 1~kpc region. This study raised questions about the building, consumption, and overall evolution of central gas reservoirs, as well as their related physical phenomena, such as gravitational torques, shear, feedback, and the exact role of bars and spiral arms in fuelling the central regions. Those questions further triggered observational surveys and numerical works to study the fuelling rate and the size of those gas reservoirs \citep{Combes2004, Garcia2005, Boone2007, Comeron2010, Sormani2019, Sormani2023B}. We still need a systematic approach to investigate numerically and characterise the different scenarios of the building and evolution of such gas reservoirs in a three-dimensional live potential (including stars, gas, and dark matter) for a set of models tailored to nearby disc galaxies. 

In this paper, we focus on the design, running, and exploitation of a suite of numerical hydro-dynamical simulations of isolated disc galaxies to study the formation and evolution of bars, and the associated building of the central gas reservoirs. We motivated such a grid of simulations using the PHANGS\footnote{Physics at High Angular resolution in Nearby GalaxieS: https://sites.google.com/view/phangs/home}-ALMA galaxy sample as a guiding baseline, which provides high-resolution, high-sensitivity, and short-spacing CO spectral cubes of nearby disc galaxies and they are ideal for comparison with the models.

In Sect.~\ref{sec:meth} we introduce the grid of models. In Sect.~\ref{sec:num} we provide details regarding the code and numerical recipes we used to perform the simulations, and in Sect.~\ref{sec:res} we present the first results extracted from that grid of simulations. We conclude in Sect.~\ref{sec:sumandconc} with a brief summary.


\section{Methods} \label{sec:meth}
\subsection{The PHANGS-ALMA sample}
As a basis to build our simulation sample, we used the PHANGS-ALMA survey \citep{Leroy2021} which consists of 118 low inclination nearby main sequence star-forming disc galaxies (in a distance range of $\sim$3-30 Mpc) with varying properties, including stellar mass, gas fraction, surface density and rotation curve \citep{Lang2020}. Those input parameters can be used to extract relevant control parameters to design a grid of initial conditions for our simulations. The most directly relevant galactic properties to build such a grid of simulations relate to the gravitational potential and associated baryonic mass distribution, hence including the stellar mass, the gas fraction and the presence of a stellar bar or not \citep{Querejeta2021}. The top panel of Fig.~\ref{fig:gf_and_histos} shows the distribution of the stellar masses in the PHANGS sample, emphasising the distribution of barred and non-barred galaxies. The stellar masses range from $\sim$10$^{9.25}$ up to $\sim$10$^{11}$ M$_{\odot}$ and the vast majority ($\sim$75 \%) of that sample shows the presence of a central bar \citep{Sophia2023}.
\subsection{The control parameters}\label{sub:cont_params}
\begin{figure*}[h]
\centering
\includegraphics[width=17cm] {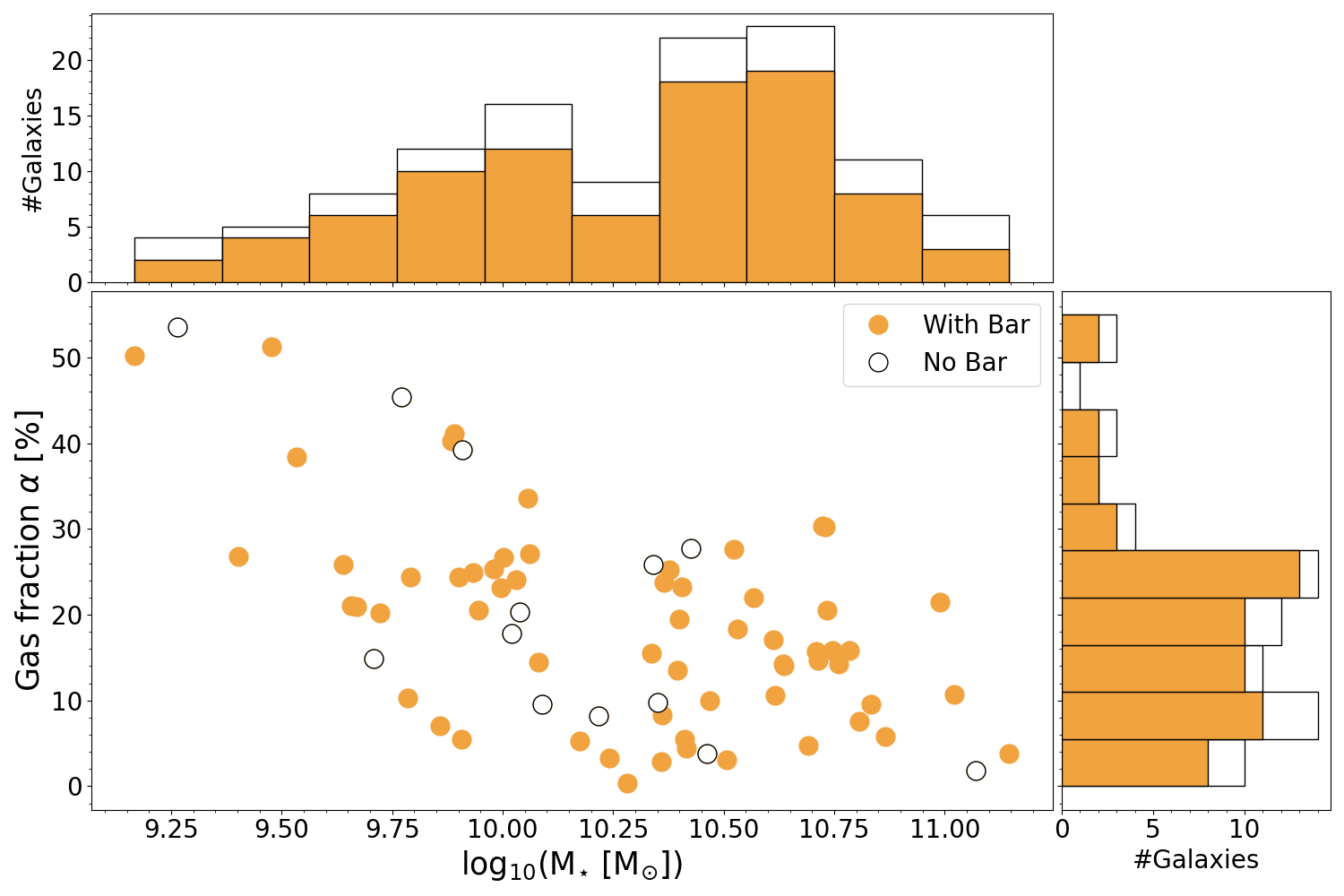}
\caption{Gas fraction ($\alpha$) as a function of the stellar mass of the galaxies in the PHANGS sample and the corresponding histograms for the stellar mass (top panel) and gas fraction (right panel) distribution (see Sect.~\ref{sub:cont_params}).}
\label{fig:gf_and_histos}
\end{figure*}

The main objective of our grid of hydro-dynamical models is to study the impact of galaxy properties on the building and evolution of the central (few 100~pc) gas reservoirs. We thus focus primarily on the properties of galaxies that we expect would more directly impact the structure of bars and spiral arms (i.e. size, strength and pattern speed). Based on the observed properties of the PHANGS sample, we focus on five galactic parameters for our models, namely:
\begin{enumerate}
    \item the stellar mass M$_{*}$;
    \item the gas fraction $\alpha = \rm{M_{g}} / (\rm{M_{g}} + \rm{M}_{\star})$ where $\rm{M_{g}} = \rm{M_{HI}} + \rm{M_{H_2}}$;
    \item the typical scale length of the stellar distribution l$_{*}$;
    \item the typical scale length of the gas (HI and H2) distribution l$_{g}$;
    \item and the central bulge mass fraction $\beta = \mbox{M}_{b}  / \mbox{M}_{\star}$.
\end{enumerate}
In this work, only four of these parameters are used as control parameters since we set the value of the gas scale length to 2 $l_{*}$.
We decompose the stellar mass into a given disc and central ellipsoid as M$_{*}$ = M$_{d}$ + M$_{b}$, with M$_{d}$ and M$_{b}$ the mass of the stellar disc and the stellar ellipsoid, respectively. In the following, we will label `bulge' this central ellipsoid, emphasising the fact that such a label is often misused to describe a central excess of light above a given larger-scale disc \citep{Gadotti2020}, leading to confusion. In our semantic usage, bulge refers to a structure that bulges out of the disc and must then be both extended (with scales significantly larger than e.g. a nucleus, or the scale height of the disc) and puffed (e.g. axis ratio significantly above 0.2). Simple relations connect the gas and stellar disc masses with the $\alpha$ and $\beta$ parameters, namely:
\begin{equation}
\label{eq:alpha}
    M_{g} = \frac{\alpha}{1-\alpha}\cdot M_{*}, \ \ \ M_{d} = (1 - \beta)\cdot M_{*}.
\end{equation}
%
%
%
We select four values for the stellar mass, namely log$_{10}$M$_{\ast}$(M$_{\odot}$) = 9.5, 10, 10.5, 11,  to cover the PHANGS range (see Fig.~\ref{fig:gf_and_histos}).

We further examine the trends of all the other four input parameters as a function of the stellar mass. The right and bottom panels of Fig.~\ref{fig:gf_and_histos} show the distribution of the gas fraction among the PHANGS galaxies and its values as a function of the stellar mass, respectively. The gas fraction tends to increase with decreasing stellar masses below $10^{10}$~M$_{\odot}$.
We assume exponential radial profiles both for the initial stellar and gas surface density ($\Sigma_\mathrm{d}$, $\Sigma_\mathrm{g}$ accordingly) profiles and fitted all azimuthally averaged observed PHANGS profiles accordingly. 
\begin{equation}
\label{eq:sig_ds}
    \Sigma_{d,*/g}(r) = \Sigma_{0,*/g} \exp \left( -\frac{r}{l_{*/g}} \right),
\end{equation}
where $l_*$ and $l_g$ are respectively the stellar and gas scale length, and $\Sigma_{0,b}$ is the initial bulge surface density. During the fitting of the stellar density profile,
we allow the simultaneous presence of a central component with a surface brightness profile described as an additional Sersic profile \citep{Sersic1963}:
\begin{equation}
\label{eq:sig_bs}
    \Sigma_{b}(r) = \Sigma_{0,b} \exp \left( - b_{n} \left[ \left( \frac{r}{R_{e,b}} \right)^{1/n_b} - 1 \right] \right),
\end{equation}
where $r$ is the bulge coordinate radius, $\Sigma_{0,i}$ is the initial surface density, R$_{e,b}$ is the scale length, $n_b$ is the Sersic index, and $b_{n}$ is a coefficient depending on $n_b$ calibrated to reach the half of the total luminosity at R$_{e,b}$. Since we wish that the three-dimensional ellipsoidal representation based on the one-dimensional Sersic function follows averaged properties of galactic bulges \citep{Sersic1963}, we forced an intrinsic ellipticity of 0.4 (axis ratio of 0.6, see \citet{Moriondo1998, Sotnikova2012}), as opposed to the standard assumptions of a spheroid (i.e. a spherical bulge). As mentioned, in the following, that component will be referred to as the `central bulge'.

\begin{figure}[h]
\resizebox{\hsize}{!}{\includegraphics{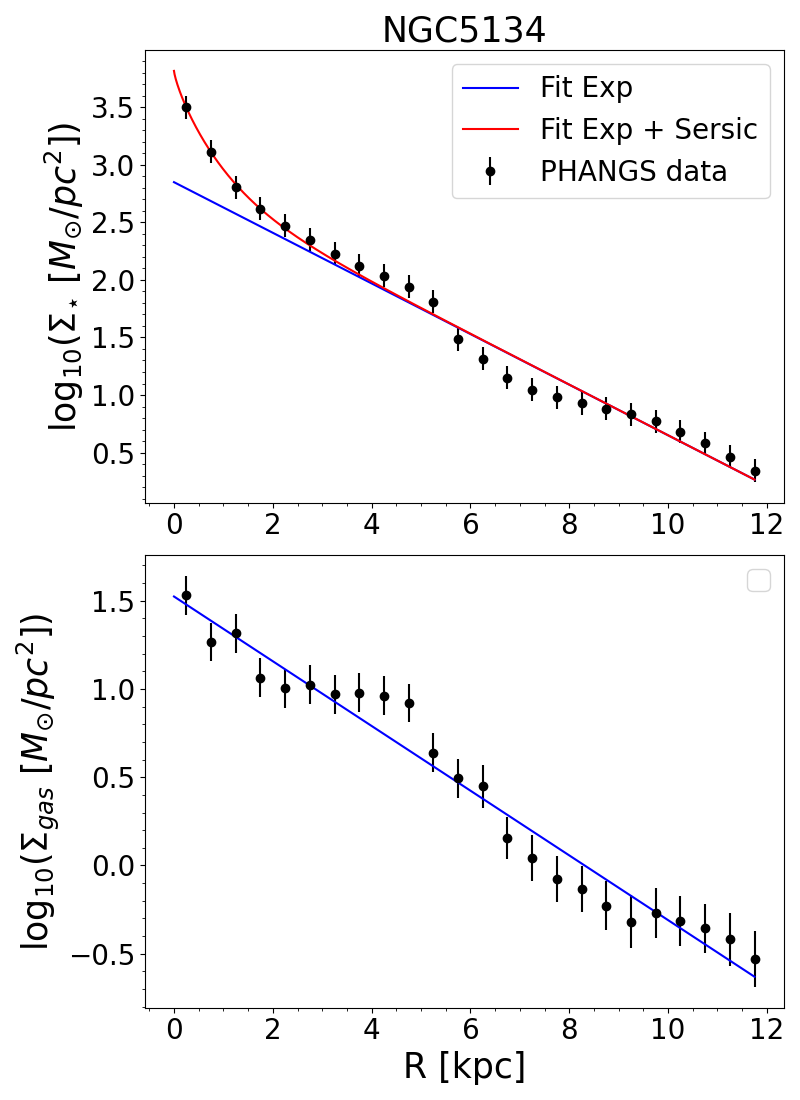}}
\caption{Illustrations of the fitted radial gas and stellar density profiles. The plots show the original data points (in black, see \citet{Sun2022}, mega-tables version 3), i.e. the azimuthal-averaged surface density profiles for the stellar (top panel) and gas (bottom panel) components of the galaxy NGC\,5134. The corresponding fits are superimposed and colour-coded accordingly (blue for the disc component, red for the total disc and ellipsoid (or bulge) component).}
\label{fig:fit}
\end{figure}

\begin{figure*}[t!]
\centering
\includegraphics[width=17cm]{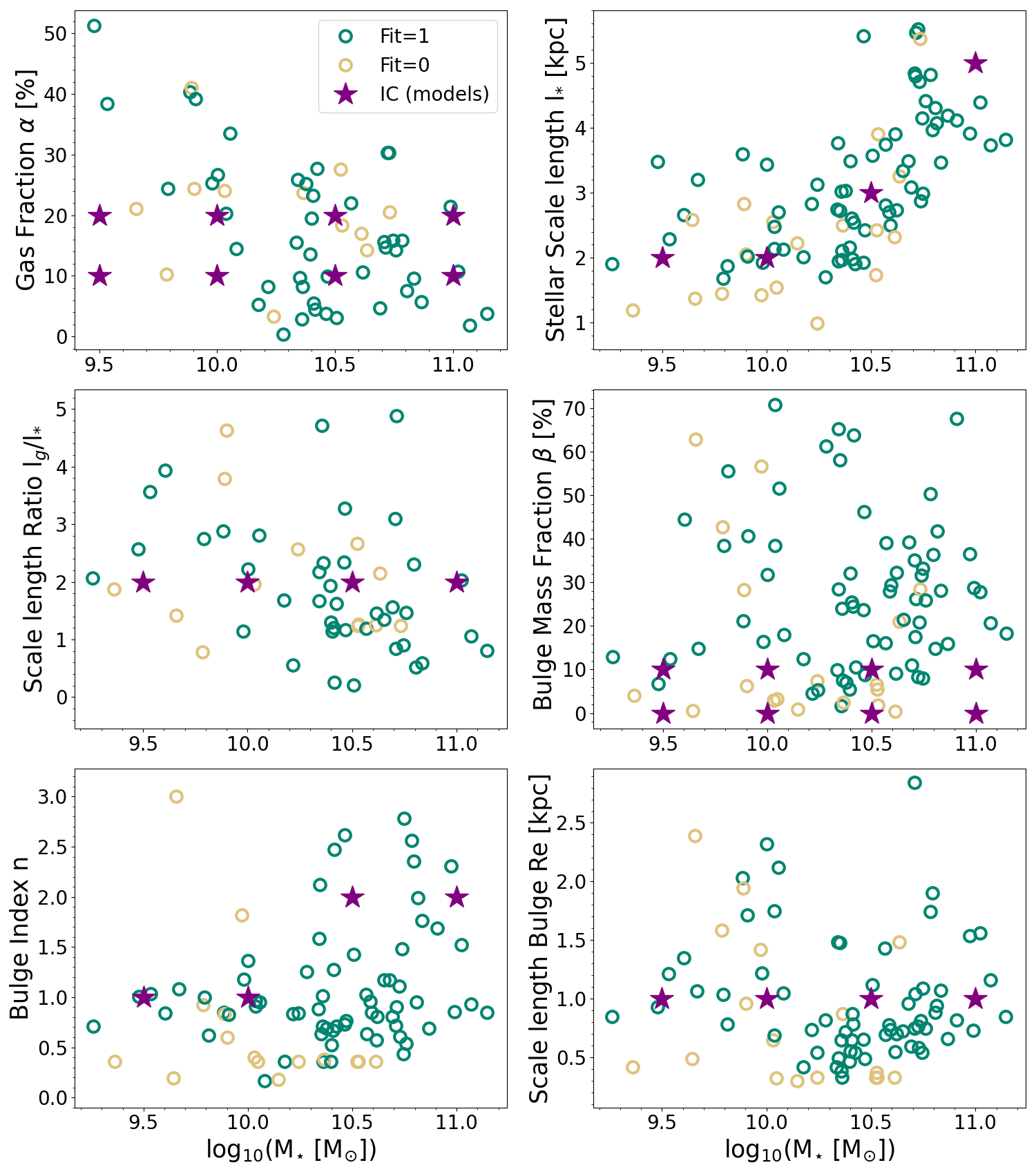}
\caption{Results of the fit for different parameters of the models as a function of the stellar mass. Top panels: gas fraction (left) and stellar scale length (right). Middle panels: scale length ratio between the gas and stars (left) and scale length of the stellar bulge (right). Bottom panels: bulge index (left) and bulge mass fraction (right). Colour circles show the actual fits (see the text in Sec.~\ref{sub:cont_params} for details) to the PHANGS-ALMA sample, while the selected values for our used initial conditions (IC(models)) are shown with purple stars.}
\label{fig:PH_IC_Tot}
\end{figure*}

Fig.~\ref{fig:fit} shows an example of the stellar (top panel) and gas (bottom panel) surface density fits for one specific PHANGS galaxy (NGC\,5134). Fits to the observed surface brightness profiles were not always satisfactory when the data could not be well represented by a single exponential or a combined exponential plus Sersic components. We thus visually classified our fits into three categories: [1] good fits, [0] reasonable fits (exhibiting radial ranges with significant residuals), and [$-1$] failed fits when no good representation of the data could be found using the above-mentioned functions (see Appendix~\ref{fig:Fit_0_1} for more details). This classification is subjective and only meant as a guide for the values of reference parameters when building up the grid of initial conditions: it does not have a significant impact on the chosen grid of models we have selected.
This analysis led us to fix a set of values for l$_{*}$, l$_{g}$, R$_{e,b}$ and $n_b$ for each galaxy in the PHANGS-ALMA sample: those are shown in  Fig.~\ref{fig:PH_IC_Tot} where they are plotted as a function of the stellar mass. We also show the values of the gas fraction $\alpha$ and those of the bulge mass fraction as derived from the obtained scale lengths.
\subsection{The grid of models}
Our grid of models relies on a selected set of values for each of the four control parameters (see previous Section). As a first approximation, this set provides a fair representation of the global trends (e.g. with respect to the stellar mass) observed in the PHANGS sample. We sometimes constrained the value of a given parameter to stay constant at all masses, at the expense of missing the more relevant range of observed parameters\footnote{Note that an extended grid of models, better covering those missed ranges, is planned (see Sect.~\ref{sec:sumandconc}).}. This is true, for instance, for the gas fractions ($\alpha$) at the lowest stellar mass bin, where we chose to only probe 0 and 10\%, thus departing from the significantly larger observed values. It is also the case for bulge mass fraction ($\beta$) for which we also kept low values (0 and 10\%) while many targets exhibit values up to 40\% or higher. This latter choice was motivated by the fact that we expect the central region to grow in mass during the secular evolution of the system and that most of the measured bulge indices are on the low side, hence most possibly reflecting the presence of a central (flattened) disc (not a puffed-up spheroid). In practice, those values have been used to build initial conditions for the hydro-dynamical simulations: they are illustrated by purple stars in Fig.~\ref{fig:PH_IC_Tot} as mentioned before and are all tabulated in Table~\ref{Tab:Grid_16}. 

In the following, all models share a common labelling scheme, namely: each model adopts a format as GxxxMxxxFxxLxBxx, where "G (stands for Galaxy)" is followed by an integer (used as an internal reference), M by the log of the stellar mass multiplied by 10, F by the gas fraction, L by the typical scale length of stars, and B by the bulge mass fraction. This represents a set of 16 high-priority models, spanning typical PHANGS parameters, and tractable in terms of computing time. This paper focuses on this preliminary first subset of 16 initial conditions, with suitable values of control parameters for the comparison between all the stellar mass bins (see Table~\ref{Tab:Grid_16}) that already covers a good range of properties for this sample. An extended grid of 54 models is planned but requires a significantly larger investment in computational time, and its analysis will be presented in a future paper.
%
%
\begin{table}[t]
\centering
\resizebox{\columnwidth}{0.42\columnwidth}{%
\begin{tabular}{|c|c|c|c|c|c|} 
\hline
Model & $\log_{10}(M_{\star}$) & $\alpha$ & $l_{\star}$ & $l_g/l_{\star}$ & $\beta$ \\
 & [M$_{\odot}$] & [\%] & [kpc] & & [\%] \\
\hline
G001M095F10L2B00 & 9.5 & 10 & 2 & 2 & 0 \\
\hline
G002M095F10L2B10 & 9.5 & 10 & 2 & 2 & 10 \\
\hline
G013M095F20L2B00 & 9.5 & 20 & 2 & 2 & 0 \\
\hline
G014M095F20L2B10 & 9.5 & 20 & 2 & 2 & 10 \\
\hline
G037M100F10L2B00 & 10 & 10 & 2 & 2 & 0\\
\hline
G038M100F10L2B10 & 10 & 10 & 2 & 2 & 10\\
\hline
G053M100F20L2B00 & 10 & 20 & 2 & 2 & 0\\
\hline
G054M100F20L2B10 & 10 & 20 & 2 & 2 & 10\\
\hline
G105M105F10L3B00 & 10.5 & 10 & 3 & 2 & 0\\
\hline
G106M105F10L3B10 & 10.5 & 10 & 3 & 2 & 10\\
\hline
G137M105F20L3B00 & 10.5 & 20 & 3 & 2 & 0\\
\hline
G138M105F20L3B10 & 10.5 & 20 & 3 & 2 & 10\\
\hline
G161M110F10L5B00 & 11 & 10 & 5 & 2 & 0\\
\hline
G162M110F10L5B10 & 11 & 10 & 5 & 2 & 10\\
\hline
G177M110F20L5B00 & 11 & 20 & 5 & 2 & 0\\
\hline
G178M110F20L5B10 & 11 & 20 & 5 & 2 & 10\\
\hline
\end{tabular}%
}
\caption{Grid containing the parameters for the initial conditions of the models of this paper. From left to right columns: label, total stellar mass, mass gas fraction, stellar scale length, scale length ratio (gas over stars) and bulge mass fraction. }
\label{Tab:Grid_16}
\end{table}


\section{Numerical simulations} \label{sec:num}
\subsection{Initial conditions}
Initial conditions for the RAMSES adaptive mesh refinement (AMR) code \citep{Teyssier2002} require information both for the particles (stars, dark matter, a central SMBH), and gas content. The grid of values gathered in Table~\ref{Tab:Grid_16} is used for the description of the distribution of each individual component. We use the Python library `pymge' based on the Multi-Gaussian Expansion \citep[MGE]{Emsellem1994, Emsellem1994_2} method to fit the associated two-dimensional distributions (e.g. exponential discs) and deproject them into three-dimensional density distributions. The vertical axis ratios of the ellipsoid resulting from that 3D deprojection are set to $q = 0.6, 0.1, 0.05$, respectively for the bulge, stellar disc and gas disc.
The dark matter is constrained by the averaged rotation curves observed via the PHANGS sample and generated using an Einasto spherical mass distribution \citep{Einasto1965, Ludlow2017} as given by:
\begin{equation}
\label{eq:rho_dm}
    \rho_{h}(r) = \rho_{h,0} \exp \left( - 2 m \left[ \left( \frac{r}{l_{h}} \right)^{1/m} - 1 \right] \right),
\end{equation}
where $\rho_{h}$ is the 3D density profile of the dark matter halo, $r$ is the spherical radius, $l_{h}$ is the scale length and $m$ is the halo index. The values we chose for the parameters of Eq.~\ref{eq:rho_dm} are summarised in Table~\ref{Tab:DM_profile}.
Once those three-dimensional profiles are fixed for all the components, we fix the number of particles and derive their initial positions and velocities by solving the Jeans equations. We assume a local anisotropy constrained by the flattening of the individual components, with $\delta = 1 - \sigma_z^2 / \sigma_R^2 = 0.6 \times (1 - q)$ where $\sigma_R,z$ the radial and vertical stellar velocity dispersion in cylindrical coordinates  \citep{Binney2009}. The more flattened the component is, the larger its (initial) velocity anisotropy. We use a constant mass resolution of $10^4$~M$_{\odot}$ for all stars present in the initial conditions (`old' stellar particles). This leads to an increasing number of stellar particles, that is, about 3$\times$10$^{5}$, 10$^{6}$, 3$\times$10$^{6}$ and 10$^{7}$, for the corresponding four stellar masses 9.5, 10, 10.5 and 11 (in log$_{10}$[M$_{\odot}$]), respectively. The number of particles for the dark matter component was fixed to 10$^{6}$ for all models.

\begin{figure*}[t]
\centering
\includegraphics[width=17cm]{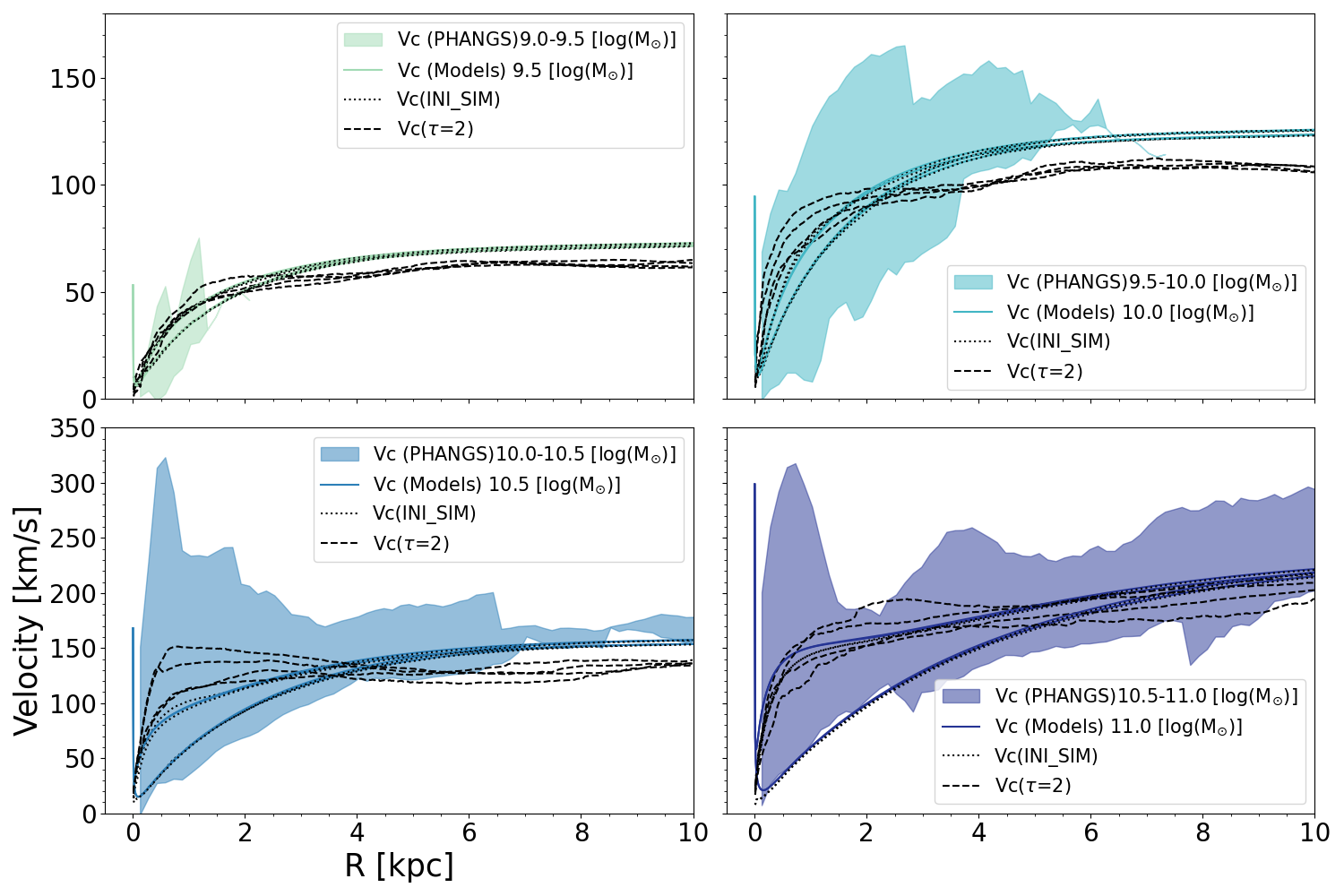}
\caption{Comparison between the observed gas velocity curves of the PHANGS sample (shaded areas) and the circular velocity profiles from our models (solid curves). The vertical lines near R=0 illustrate the contribution of the central SMBH. The dotted curves represent the circular velocities extracted from the first snapshot (INI\_SIM) of each simulation and are consistent with the analytic derivation associated with the initial conditions. The dashed curves represent the circular velocities at $\tau=2$ (see Sect.~\ref{subsec:phases}, which corresponds to twice the bar formation time of our simulated galaxies). As mentioned in the text, the PHANGS rotation curves are meant as guidelines to build the initial mass models.}
\label{fig:IniVc}
\end{figure*}

Fig.~\ref{fig:IniVc} shows a comparison between the PHANGS rotation curves (shaded areas, encompassing individual velocity profiles for a given stellar mass bin, \citealt{Lang2020}) and our 16 models (solid curves) for the different mass ranges. Our models are, by design, in good agreement with the PHANGS galaxies, reproducing the global trend of real galaxies. We note that the limited extent in the observed rotation curves for the lower mass bins is expected considering the smaller size and shorter radial extent of the (mostly CO and HI) tracers \citep{Leroy2019}. We also emphasise that those circular velocity curves represent the initial state of our hydro-dynamical simulations, and are thus meant to evolve with time (i.e. see the black curves). It is worth mentioning that the PHANGS rotation curves are the observed gas velocity profiles. Converting such velocity profiles to actual circular velocities, as measured in our simulations, would require detailed modelling, including, for example, the effect of asymmetric drift and non-circular motions. Considering the relatively low velocity dispersion of the molecular gas, we do not expect the impact of asymmetric drift to be significant. Non-circular motions may, however, dominate in the central region. This motivated us to use those PHANGS observed rotation curves only as guidelines to build our mass models.
\begin{table}[h]
\centering
\begin{tabular} {|c|c|c|c|c|} 
\hline
$M_{\star}$ & $\rho_{h,0}$ & $l_{h,0}$ & $m$ & $M_{BH}$ \\
$\log_{10}$([M$_{\odot}$]) & [M$_{\odot}$/pc$^{3}$] & [kpc] & & $\log_{10}$([M$_{\odot}$]) \\
\hline
9.5  & $3 \times 10^{-4}$ & 20 & 3 & 5.5 \\
\hline
10 & $6 \times 10^{-4}$ & 25 & 3.5 & 6 \\
\hline
10.5 & $7 \times 10^{-4}$ & 30 & 1.5 & 6.5 \\
\hline
11 & $8.5 \times 10^{-4}$ & 35 & 2 & 7 \\
\hline
\end{tabular}
\caption{Grid of values for the parameters describing the dark matter profiles (i.e. $\rho_{h,0}$, the halo density; $l_{h,0}$, the typical halo scale length; and $m$, the halo index) as a function of the stellar mass $M_{\star}$. Those values have been inspired by the work of \citet{Kun2017}, but their final values have been chosen to match the PHANGS rotation curves. The central black hole masses $M_{BH}$ we use in the initial conditions is also shown in the last column.}
\label{Tab:DM_profile}
\end{table}
\subsection{Refinement strategy}
\label{sect:refine}
All simulations in the three lower stellar mass bins (resp. higher mass bin) are run within a 100~kpc (resp. 200~kpc) cubic box. A refinement level $l$ thus corresponds to a 100/2$^{l}$ (resp. 100/2$^{l}$)~kpc cubic cell. We imposed a minimal refinement level of 7 (resp 8; $\sim 780$~pc) and a maximal one of 13 (resp. 14; $\sim 12$~pc). The refinement strategy we adopt for the gas is based on a Jeans polytropic approximation \citep[see][]{Renaud2013} allowing us to set a realistic threshold for the amount of gas contained inside a cell. The maximum mass of gas contained inside a cell at the second last level (l=12 or 13, 24~pc) before the triggering of the refinement to the last level (l=13 or 14, 12~pc) is around 14000~\Msun. An additional refinement criterion is constrained by the minimum number of particles per cell (stars and dark matter) that is set to 8. 
\subsection{Numerical recipes}
A number of standard physical processes were implemented including gas cooling, star formation, stellar feedback and the evolution of metals such as iron and oxygen (see \citealt{Agertz2013} for more detailed information). The physics of the heating follows a uniform UV background model proposed by \citet{Haardt1996} and calibrated for redshift $z=0$. The physics of the cooling is implemented and divided into two different regimes. The first regime goes down two $10^{4}$ K and uses collisional de-excitation and atomic recombinations. The second regime, below $10^{4}$ K, follows the table of \citet{Sutherland1993}.
The formation of new stars is triggered above a gas density threshold via a constant efficiency, as given by the following equation:
\begin{equation}
\label{eq:sfr}
      \dot{\rho}_{*} = \frac{\rho_{g}}{t_{SF}} \ \text{for} \ \rho_{g} \ge m_{h} n_{*}  ,
\end{equation}
where $\dot{\rho_{*}}$ is the star-formation rate (SFR), $\rho_{g}$ is the gas density, $m_{h}$ is the mass of the hydrogen atom and $t_{SF}$ is the typical gas depletion time ($\sim$ 2 Gyr). The latter is given by:
\begin{equation}
\label{eq:t_sf}
      t_{SF} = t_{ff}/\epsilon_{ff},
\end{equation}
with $t_{ff}$ the local free fall time and $\epsilon_{ff}$ the star formation efficiency per free-fall time. We set the efficiency to 2\%, and $n_{*}$ to 100 cm$^{-3}$, a posteriori checking that our systems follow the trend observed in the PHANGS galaxies (see Fig.~\ref{fig:SFR}). Stellar feedback, which releases energy, momentum and metals in the surrounding ISM, is implemented by taking into account the contribution from supernovae (SN) SNIa, SNII and stellar winds. The energy released by the supernova depends on the local cooling radius \citep[see][]{Kim2015} Finally, we also adopt solar metallicity as initial condition (Z = 1~Z$_{\odot}$), which does not have a significant impact on the gas processes (star formation, feedback).

\section{Results} \label{sec:res}
In this section, we first describe some general properties of the evolved systems and then illustrate the global time evolution and bar formation. We briefly discuss the bar formation times in the light of past results, and consider the time evolution in terms of specific successive phases. We finally hint at a significant difference in the phases of the building of a central gas concentration impacting the way star formation proceeds in the lowest mass bin of our sample.

\subsection{General properties and time evolution}

\subsubsection{The star-forming main sequence}
\label{sec:ms}
\begin{figure}[t!]
\resizebox{\hsize}{!}{\includegraphics{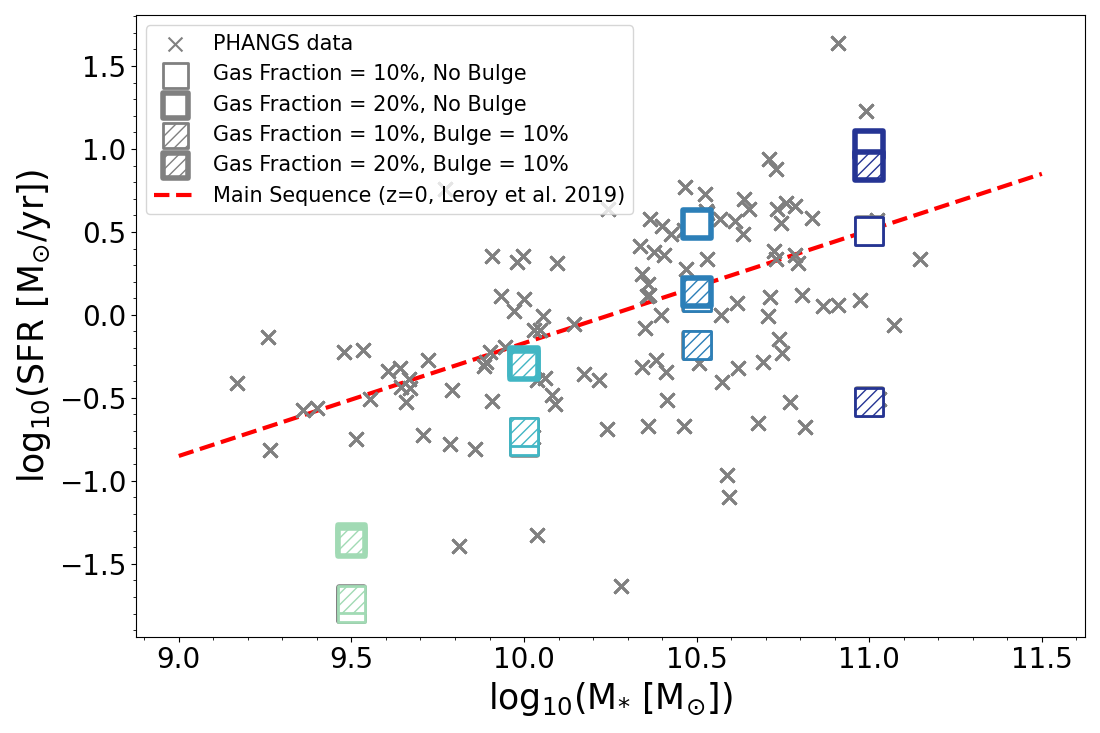}}
\caption{Comparison of the SFR stemming from the PHANGS data sample (grey crosses) and the SFR computed from the simulated galaxies (coloured squared). The colour of the squares corresponds to the different stellar masses. The red dashed line represents the star formation main sequence at $z=0$ from \citet{Leroy2019}.}
\label{fig:SFR}
\end{figure}

We want our simulated galaxies to be representative of the star-forming main sequence of nearby disc galaxies, hence located on the Kennicutt-Schmidt relation \citep{Schmidt1959, Kennicutt2007}. Our choices for the initial properties (e.g.distribution of baryons, gas fractions, etc.) and sub-grid recipes (e.g. star formation efficiency, see Sect.~\ref{sect:refine}) do not guarantee such a result, because star formation may partly regulate itself in such numerical experiments \citep{Ostriker2022}. We thus computed the global SFR for all 16 simulations and compared its characteristics and trends with the instantaneous SFR of the PHANGS-ALMA sample as shown in Fig. \ref{fig:SFR}. We computed the SFR by taking the time average of the SFR, only including times between $\tau=1$ and $\tau=2$ (see Sect.~\ref{subsec:phases}, which corresponds to the starburst phase of our simulated galaxies), thus focusing on the active star formation regime. 

In Fig.~\ref{fig:SFR}, we confirm that our simulations are almost all following the PHANGS main sequence, except for galaxies with the lowest stellar mass bin (i.e. 9.5 in log$_{10}$(M$_{\odot}$)): those have SFR lower than the observed values in this sample. This is expected since we intentionally chose to run this first subset of simulations with only two values of the gas mass fraction, those being significant underestimating the observed gas fractions ($\sim$30-40\%) for that stellar mass bin, as shown in Fig.~\ref{fig:PH_IC_Tot}. Higher gas fractions (thicker squares) naturally lead to a higher SFR, and models with a bulge (hatched squares) tend to have lower resulting SFR as expected if the self-gravity of the disc is lowered. The impact of the bulge reaches its maximum for the bulged model with the highest stellar mass, and a gas fraction of 10\%. We describe in more detail the characteristics of this model in the next sections.
\subsubsection{Building of the bar}
\label{subsec:building_bar}
\begin{figure*}[t!]
\centering
\includegraphics[width=18cm]{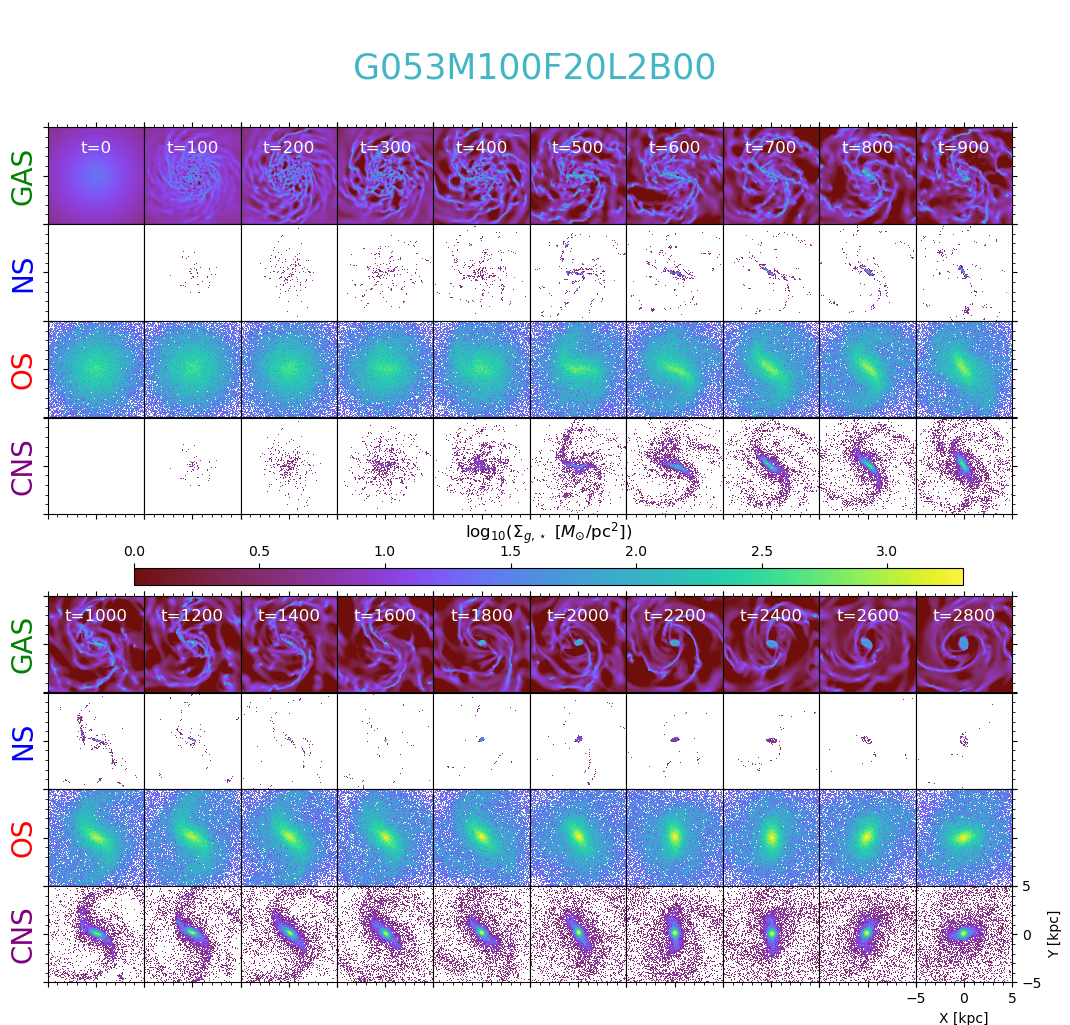}
\caption{Evolution over time (in millions of years) of the gas and stars of the model labelled G053M100F20L2B00 (one of the sixteen simulations presented in this paper). The two big panels illustrate four main stages, chronologically ordered from top left (0-400 Myr), to top right (500-900 Myr), bottom left (1000-1800 Myr) and bottom right (2000-2800 Myr). \textbf{Top left:} first spiral structures, cooling of the gas, onset of star formation, initial bar structure emerging. \textbf{Top right:} bar strengthening and active local star formation. \textbf{Bottom left:} building of a central concentration of gas (and new stars). \textbf{Bottom right:} growth of the gas reservoir. In each big panel, from top to bottom, we present maps for the gas mass density (\textcolor{OliveGreen}{GAS}), the mass of young stars ($\leqslant$ 50 Myr, \textcolor{blue}{NS}), of old stars (\textcolor{red}{OS}) and finally the cumulative mass of new stars (formed since the beginning of the simulation, \textcolor{violet}{CNS}). Each panel shows a $10\times10$~kpc$^2$ region and the colour scaling is adapted for each panel.}
\label{fig:E_G053}
\end{figure*}

Figure.~\ref{fig:E_G053} illustrates the global evolution of one galaxy over time by showing the maps of gas, new and old stars. We start with an axisymmetric distribution of gas and stars, which rapidly develops low-contrast spiral arm structures. Gas follows up by cooling and forming new stars (top left panel). After a few hundred Myr we start to observe the formation of a bar and spiral arms in the old and new stars (top right panel). Once the bar has formed, the gas concentration in the central region starts to increase, leading to the build-up of an early gas reservoir (bottom left panel). This could be induced by the emergence of an inner Lindblad resonance \citep[ILR:][]{Lin2008, Sormani2023}. Finally, we observe the building and growth of a central gas reservoir (bottom right panel). 

This scenario is similar for most of the 16 simulated galaxies and the evolution over time of the 16 simulations is shown in the Appendix~\ref{app:Evol_tau} where the different times are chosen to emphasise the evolution of the building of the gas reservoir within the central 1~kpc region (see Sect.~\ref{subsec:central_kpc}).

\begin{figure}[h!]
\resizebox{\hsize}{!}{\includegraphics{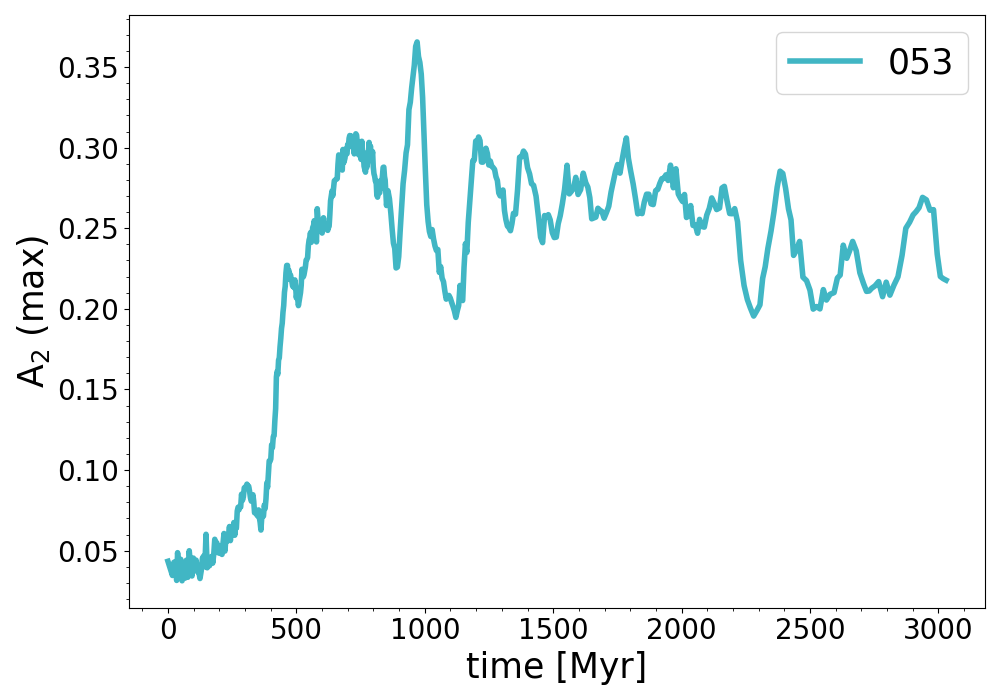}}
\caption{Evolution of the bar strength (measured through the \A\ Fourier coefficient) for the model G053.}
\label{fig:a2_G053}
\end{figure}

One of the main features developed by almost all our simulations is a stellar bar (except for models G162M110F10L5B10 and G178M110F20L5B10). We present model G053M100F20L2B00 in Fig.~\ref{fig:E_G053} to illustrate the growth of the bar witnessed in our simulations (see third row). To more quantitatively characterise the evolution of the bar over time, we estimated the amplitude of the bar via a polar ($R, \theta$) Fourier decomposition of the face-on surface stellar mass density, and specifically used the traditional \A\ Fourier coefficient \citep{Efstatiou1982, Atha2002, Atha_2_2013} given by
\begin{equation}
     A_2 =\frac{\sqrt{a_2^2 + b_2^2}}{a_0}, \ a_n = \sum_{i=1}^{N} m_i \cos(n\theta_i), \ 
     b_n =\sum_{i=1}^{N} m_i \sin(n\theta_i),
\end{equation}
where m$_{i}$ is the mass of the $i$-th star and $\theta_{i}$ its corresponding position angle. The full radial \A(R)\ profile is then used, for example, to track its maximum over time, giving us a quantitative assessment of the evolution of the bar strength, as shown in Fig.~\ref{fig:a2_G053} for the model labelled G053M100F20L2B00. In that Figure, we see that the bar grows rapidly during the first 500-600 Myr, roughly following an exponential behaviour \citep{Binney2020, Bland2023}. Then, it reaches a first maximum after $\sim 750$~Myr and a second after $\sim 1000$~Myr. This characteristic shape of \A\ was already reproduced by \citet{Atha_2_2002}. At later times, \A\ exhibits some oscillations until the end of the simulation. These modulations coincide with the relative cycling (and alignment) between the bar and the outer spiral arms which have different pattern speeds \citep{Hilmi2020}.

We emphasise that the resulting \A\ values should depend on the tracer to compute it, for instance, using particle masses, accounting only for old or newly formed stars, or using luminosity-weighted quantities as is naturally the case for observations. It may also be influenced by the presence of dust and by additional processing stages (e.g. deprojection of observed photometry). 
Still, the \A\ values found in our simulations are in the range $[0.2 - 0.5]$, consistent with the ones observed by \cite{Diaz2016} for star-forming late-type galaxies, and also consistent with the majority of barred galaxies in the PHANGS sample. \citet{Sophia2023} reports significantly larger values (\A~$> 0.4$), but those are representative of the high tail of the bar strength distribution. The fact that we do not reproduce this tail of high \A\ values may naturally arise from the early bar phase that we are probing with our simulations (i.e. up to about 3~Gyr). Some of the strong bars observed in the PHANGS sample could be at a later evolutionary stage for which the bar has experienced a second (secular) growth \citep[see][]{Atha2013}.

\subsubsection{The central 1~kpc} \label{subsec:central_kpc}
\begin{figure*}[h!]
\centering
\includegraphics[width=17cm]{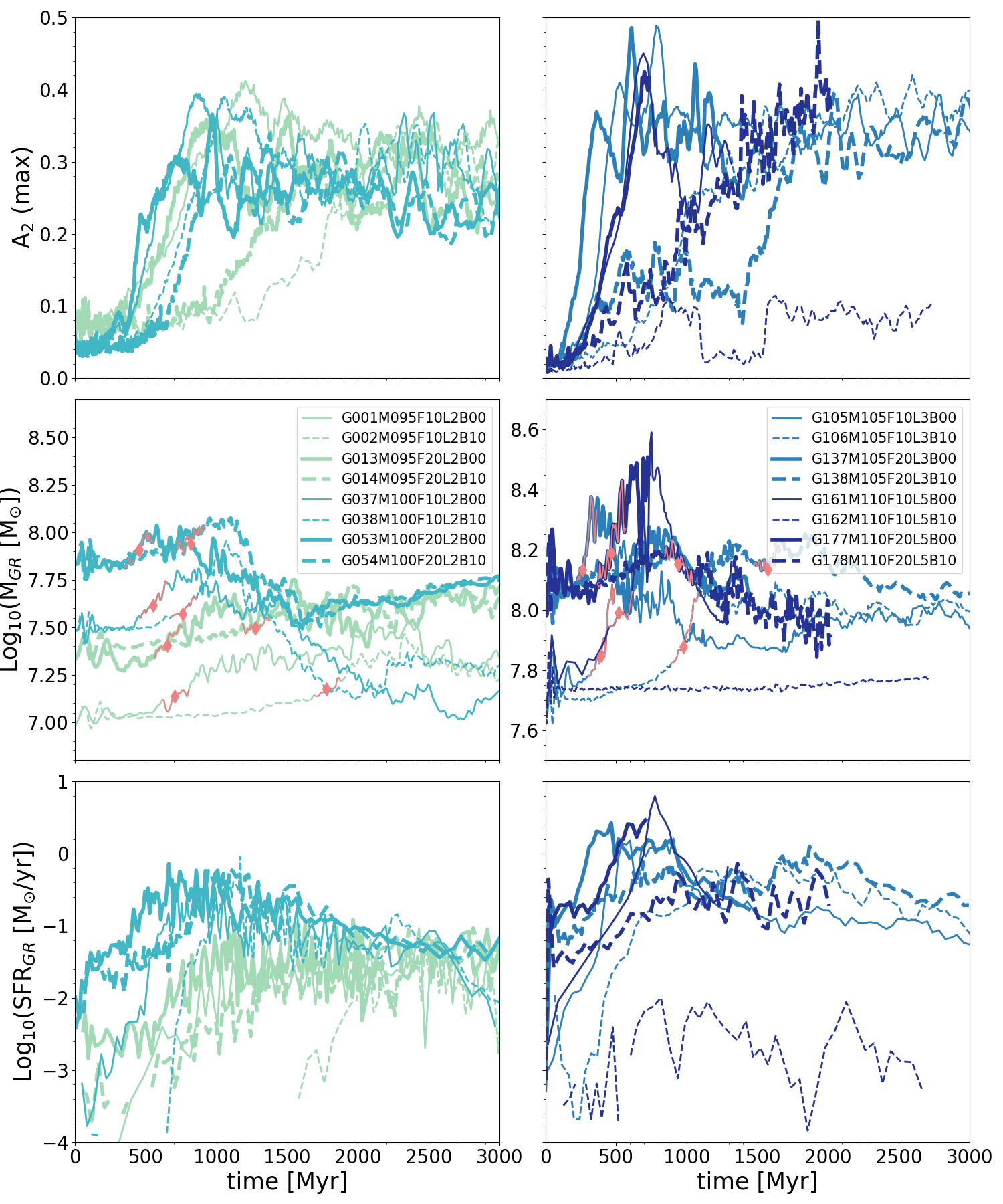}
\caption{Evolution of the maximum of the \A\ coefficient (top panels), the mass of gas within 1 kpc (middle panels), and the SFR (bottom panels) over time for the 16 simulations (evolved until max 3 Gyr). The colour gradient represents the four stellar masses (9.5, 10, 10.5 and 11 [log(M$_{\odot}$)]) from the lightest to the most massive. The left column represents the less massive galaxies (9.5 and 10, in green and cyan, respectively) and the right column shows the most massive ones (10.5 and 11, in blue and dark blue, respectively). The solid and dashed curves illustrate the models without and with a bulge, respectively and the thickness of the lines accounts for the gas fraction (10 and 20\% for the thinnest and thickest, respectively). The typical bar formation time \tb\ is also shown (red diamonds) within a time interval of 200 Myr (red part of the curves).}
\label{fig:GR_BAR_SFR}
\end{figure*}
In the PHANGS-ALMA galaxy sample,  we observe inner molecular rings with typical sizes ranging from $\sim 100$~pc to $\sim 1$~kpc \citep{Leroy2021}. Our simulations suggest their size and mass evolve with time. We thus monitored the evolution of the central gas mass concentration and star formation within the central 1~kpc, and their relation with the bar using our 16 simulations. In the rest of this section, we refer to the (central 1~kpc) `gas reservoir' to describe the mass of gas inside the central 1~kpc region. 

In Fig.~\ref{fig:GR_BAR_SFR}, the left (resp. right) column presents time evolution profiles for models with the lower (9.5 and 10 in log$_{10}$(M$_{\odot}$)) (resp. higher, 10.5 and 11 in log$_{10}$(M$_{\odot}$)) stellar masses. The middle panels show the evolution of the total gas mass contained inside a cylindrical radius of 1~kpc and a thickness of 1~kpc. The initial gas mass within 1~kpc correlates with the initial stellar mass and gas fraction as expected. After a generic plateau lasting about 300~Myr, the gas concentration increases steadily. This increase of the gas mass inside the central 1~kpc region is coincident with an increase of the SFR: those roughly correspond to \tb, the time at which the \A\ coefficient reaches the value of $\sim 0.2$, whatever the stellar mass, gas fraction or the presence of a bulge. Since bulges tend to increase \tb\ (see Sections~\ref{sec:ms} and \ref{sec:formtime}), models with bulges also show a corresponding delay in the start of the increase of the gas mass.

G162M110F10L5B10 seems to be a peculiar case among the two models that do not form a bar: \A\ never reaches a value of 0.2. Gas is consequently not funnelled towards the centre as for other models, and there are no star formation bursts with SFR staying around $10^{-3}$~\Msun/yr. In that simulation, the bulge has a stabilising effect, enough in this specific case to prevent any bar to form for the duration of our simulation. This is a well-known result, and has been already studied and recently emphasised in various papers \citep[see e.g.][]{Sellwood2001, Saha2013, Kataria2018, Fujii2019}. In that context, the bar in G162M110F10L5B10 possibly has a formation time \tb\ significantly larger than a Hubble time \citep[see also][]{Bland2023}.
\subsection{Typical bar formation time}
\label{sec:formtime}
\begin{figure}[h!]
\resizebox{\hsize}{!}{\includegraphics{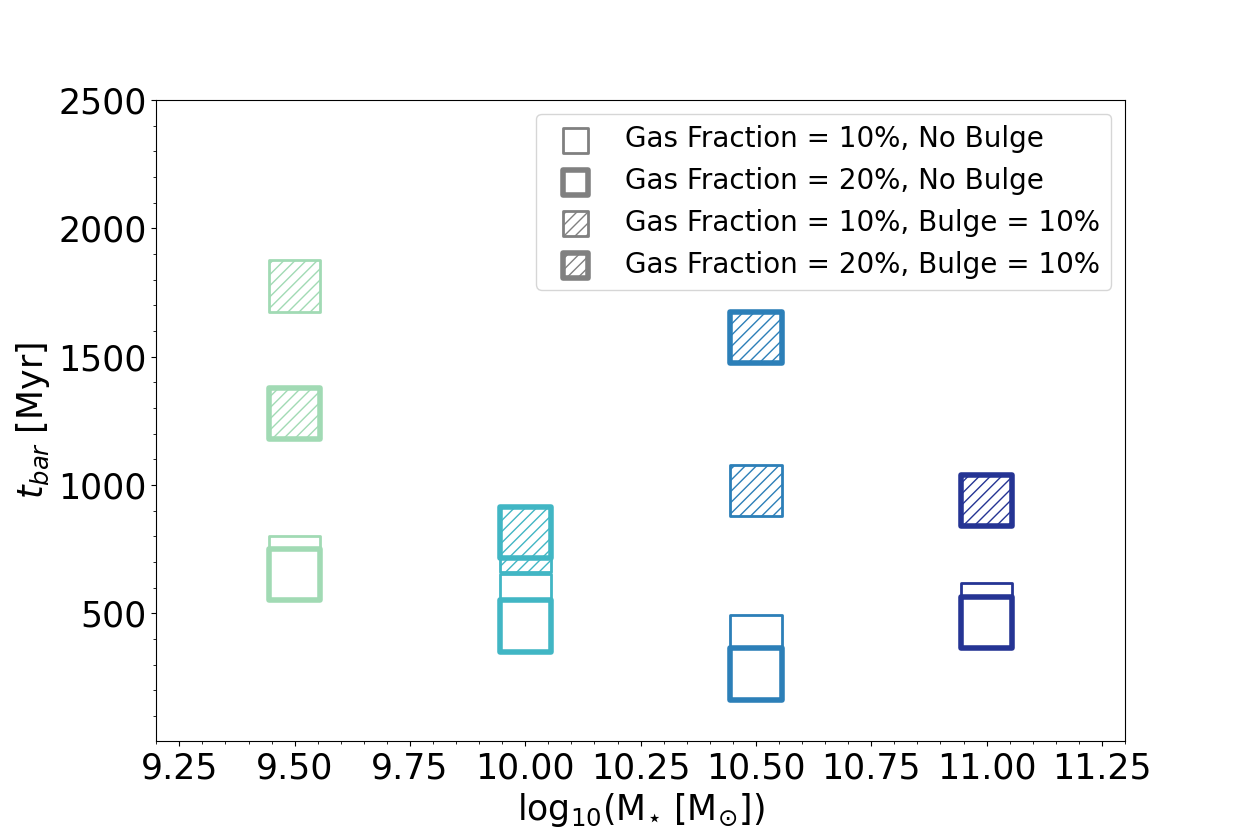}}
\caption{Typical bar formation time \tb\ as a function of the stellar mass for all 16 simulations. The colour of the squares corresponds to the different stellar masses. The size of each square represents 200~Myr (i.e. $\pm 100$~Myr), hence illustrating the uncertainty in \tb\ (see Sect.~\ref{sec:formtime}).}
\label{fig:ta2}
\end{figure}
\begin{figure}[h!]
\resizebox{\hsize}{!}{\includegraphics{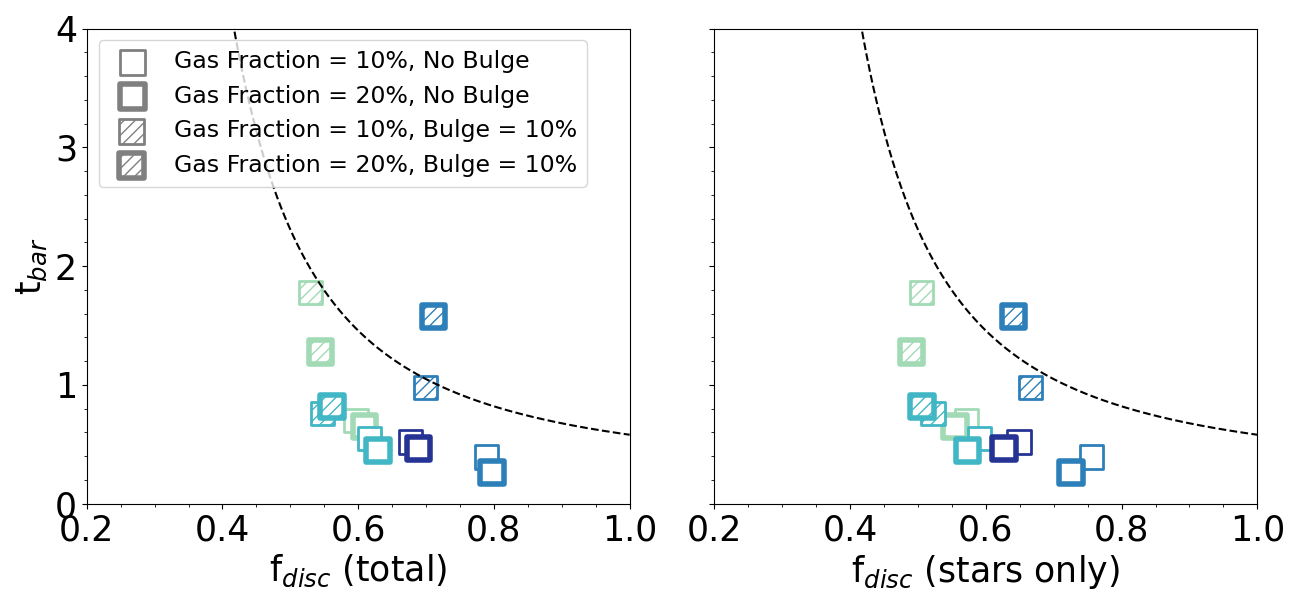}}
\caption{Typical bar formation time \tb \ as a function of f$_{disc}$ as defined in \citet{Bland2023}. The dashed lines show the relation given by \citep{Fujii2019}. $f_{disc}$ is derived using the sole stellar disc (right panel) or the total baryonic content (i.e. stars and gas; left panel). In the right panel, $f_{disc}$~(stars only) decreases when the gas fraction increases as expected (the relative contribution of the stellar disc gets smaller), while in the left panel, $f_{disc}$~(total) increases with the gas fraction (as the total baryonic contribution increases). As in Figure~\ref{fig:ta2}, the size of each square represents 200~Myr (i.e. $\pm 100$~Myr), hence illustrating the uncertainty in \tb\ (see Sect.~\ref{sec:formtime}).}
\label{fig:ta2_fdisc}
\end{figure}
Using simulation G053M100F20L2B00 as an example, we witness the onset of bar formation as early as $t=300$~Myr (Fig.~\ref{fig:E_G053}). However, the bar seems to more robustly emerge between t=400 and t=500~Myr and this corresponds with the time when \A\ reaches a value of $\sim 0.2$. This is also true for the other simulations in our sample. In the following, we therefore use \tb, the time corresponding to \A$=0.2$, as the reference time for the bar formation (see Sect.~\ref{subsec:central_kpc}). Such a reference measurement is commonly used in the literature \citep{Atha2002,Fujii2018}.

All simulations, except two (models G162M110F10L5B10 and G178M110F20L5B10) form bars. The values for \tb\ are shown for the 14 simulations forming a bar in Fig.~\ref{fig:ta2}. To further probe the stability of our models against bar formation, we have applied the criterion established by \citet{Efstathiou1982} for all 16 models. We indeed find that only the two above-mentioned models do overshoot the threshold for stability.
While this is an interesting result, suggesting that we should indeed not expect the formation of a bar in models  G162M110F10L5B10 and G178M110F20L5B10, we caution the reader regarding the use of that stability criterion as a robust quantitative diagnostic. As pointed out by \citet{Atha2008}, this criterion does not take into account the interaction between the disc and the halo, nor the velocity dispersion, which plays a major role in disc stability. They also suggested that this parameter is not designed for the stability analysis of multi-component discs. This has been also shown in a recent work carried out by \citet[][]{Romeo2023}, whose conclusion is that this parameter fails at separating barred and non-barred galaxies in 55 per cent of the cases.

We further emphasise the fact that \tb\ should be considered a relative indicator that exhibits significant systematics. We have conducted a series of tests using the simulation G037M100F10L2B00 as a reference, varying the initial number of particles and the maximum level of refinement, and even shutting off the gas cooling and star formation recipes. As long as we keep the gas disc live within the simulation, we do not witness a significant change in the evolution of the \A\ profile and thus in \tb. When the number of disc particles is changed (from 0.5 to 1, 2, 5 or 10 million), the time increase (i.e. increasing rate) of \A\ looks similar. However, the starting time of that initial increase, tracing the growth of $m=2$ modes typically varies from one simulation to the next by $\pm 75$~Myr. That timing offset does not seem to depend monotonically on the number of particles, as it is, for instance, slightly lower for 2 and 10 million disc particles, and thus higher for 1 and 5 million disc particles. We interpret it as partly due to the contribution of the spiral modes that may interfere with the bar mode, as well as variations in the initial relaxation phase. While it is beyond the scope of this study to further probe such a dependence, it does show that a significant systematic uncertainty should be included when discussing \tb\ measured in such a way. In the rest of the paper, we added a (conservative) systematic uncertainty of $\pm 100$~Myr on our \tb\ measurements and checked that it did not impact our results.

The most obvious trend associated with \tb\ is connected with the impact of the initial ellipsoid (or bulge). When an ellipsoid or bulge (hatched squares in the figure) is added to the model, a delay of a few 100's of~Myr in the bar formation is induced, leading to a delay in the building and fuelling of the central 1 kpc (see also Sect.~\ref{subsec:central_kpc}). Note that the induced time delay is significant, and much larger than the above-mentioned systematic uncertainty in \tb. For models without bulges (empty squares), a larger gas fraction (thick squares) tends to decrease \tb\, confirming early results by \citet{Atha_2_2013} who found a similar outcome for systems with up to 50\% of gas. For more gas-rich systems (e.g. gas fractions equal or larger than 75\%) \citet{Atha_2_2013} witnessed an inverse trend considering the initial growth of the bar. They also find that such trends significantly depend on the halo triaxiality (see their Fig.~7), something we do not test with our present simulations. We note that the delay in the early bar growth due to higher gas fractions in our simulations does not simply hold for our bulged models, suggesting, together with the above-mentioned results from \citet{Atha_2_2013} that a more relevant parameter may drive the initial bar growth.

\citet{Fujii2018} recently suggested to use of f$_{disc}$, the ratio between the disc mass and the total galaxy mass within $2.2\, \mbox{R}_{disc}$ as a relevant driving parameter for the bar growth timescale. This parameter f$_{disc}$ is admittedly one way to quantify the importance of the stellar disc within the global potential (baryonic + dark matter) that may be connected with the onset of bar instabilities. \citet{Fujii2018} used a series of pure N-body simulations of systems with stellar masses of a few $10^{10}$~\Msun, with varying bulge to disc ratios, using 500 million particles (8 million for the baryonic disc) and 10~pc softening length to study timescales associated with bar and spiral growth \citep[see also][]{Valencia2017}. They suggested a relation between f$_{disc}$ and \tb. \citet{Bland2023} reviewed the \citet{Fujii2018} results by running simulations both using dry runs (stars and dark matter particles only) and a few wet runs (including gas but excluding cooling and star formation) using the AMR code RAMSES, with spatial sampling of 6 and 12~pc: they mostly confirmed the existence of a relation between f$_{disc}$ and \tb, the 'Fujii relation', albeit with a large scatter for f$_{disc}$ larger than 0.5 (their Fig.~4).
\citet{Bland2023} further suggested that the addition of gas tends to decrease the value of \tb, as also witnessed in our bulgeless simulations \citep[but not in our bulged models; and see][]{Atha_2_2013}.

We take the opportunity of our sample of simulations to test further the validity range of the 'Fujii relation'. 
Our simulations have similar resolution and particle numbers, and use a very similar code than \citet{Bland2023} but include cooling and star formation. As compared to the original \citet{Fujii2018} simulations that are pure N-body via a tree-code \citep{BarnesHut86}, our simulations have 8 times fewer particles in the disc (and a mass resolution for the dark matter halo about 100 times larger) and again include gas and star formation. Figure.~\ref{fig:ta2_fdisc} specifically illustrates the trend of \tb\ as a function of f$_{disc}$ for our set of simulations, emphasising the fact that we are so far only probing values of f$_{disc}$ larger than 0.5 by design.

The left (resp., right) panel shows the \citet{Fujii2018} relation for f$_{disc}$ (dashed lines) computed with the star and gas (resp. only star) component. The inclusion of the gas as part of the disc potential (left panel) increases f$_{disc}$ only slightly, as expected: for galaxies without a bulge, increasing the gas fraction also systematically decreases \tb: that can be understood in terms of the destabilising effect of the additional disc of gas. For the more massive simulations with bulges, the trend seems the opposite, illustrating that the addition of an ellipsoid dominates the budget for \tb. Finally, the simulation at the lowest stellar mass bin and with a bulge shows a smaller \tb\ when the gas fraction is increased. 

At fixed f$_{disc}$, our sample of simulations seems to overall lie a factor of two below the \citet{Fujii2019} relation, with differences of about 500~Myr or more, well beyond the expected uncertainty associated with the measured \tb\ values. As emphasised, one difference with the results from \citet{Fujii2019} is that our simulations contain gas. While, as mentioned above, the bar formation time may depend on the gas fraction, this discrepancy does not change significantly if we include the gas component in the value of f$_{disc}$. Another significant difference between our simulations and those by \citet{Fujii2019} is that the latter used pure N-body simulations with a large number of particles (with a softening length of about 10~pc). Still, we are in a regime where we should have enough mass resolution, at least for the 3 larger mass bins (leading to a minimum of 2~million particles), and we do not detect a trend in stellar mass in the context of the \tb\ discrepancy with \citet{Fujii2019}. Even considering the added systematic error of $\pm 100$~Myr (see the beginning of the present Section), we do not retrieve the \citet{Fujii2018} relation. We also need to emphasise the fact that simulations performed by \citet{Bland2023} use a similar setup as ours (and the same AMR code, RAMSES), and do not see such a discrepancy with \citet[][their Fig.~4]{Fujii2019}. 

More fundamentally, other differences could be at the root of this discrepancy, including more concentrated discs and dark matter distributions in \citet{Bland2023}, different initial velocity distributions (anisotropies), or the lack of cooling, star formation and feedback in the latter experiments. As emphasised by \citet{Atha_2_2013}, other parameters, including the shape of the halo and the initial dynamical set up for the disc, can strongly influence the growth of unstable modes, including the bar and spirals.

Our results confirm and extend previous studies showing that \tb\ seems to depend on many details in the initial conditions, including the steepness of the inner potential (e.g, via the index of the Sersic bulge), the overall contribution of the disc components \citep[w.r.t. other baryonic and non-baryonic components; see also][]{Atha2013}. This may naively suggest that \tb\ is not fully described by a single parameter such as f$_{disc}$. However, we cannot yet conclude, as we would need to probe a more extended set of simulations, including, for example, models with varying properties (e.g. anisotropies) at fixed f$_{disc}$, and models with value of f$_{disc}$ below 0.5 to detect the exponential increase of \tb\ (and where values of \tb\ are significantly larger than about 2~Gyr).

\subsection{Evolutionary phases of isolated barred discs} \label{subsec:phases}
\begin{figure*}[h!]
\centering
\includegraphics[width=17cm]{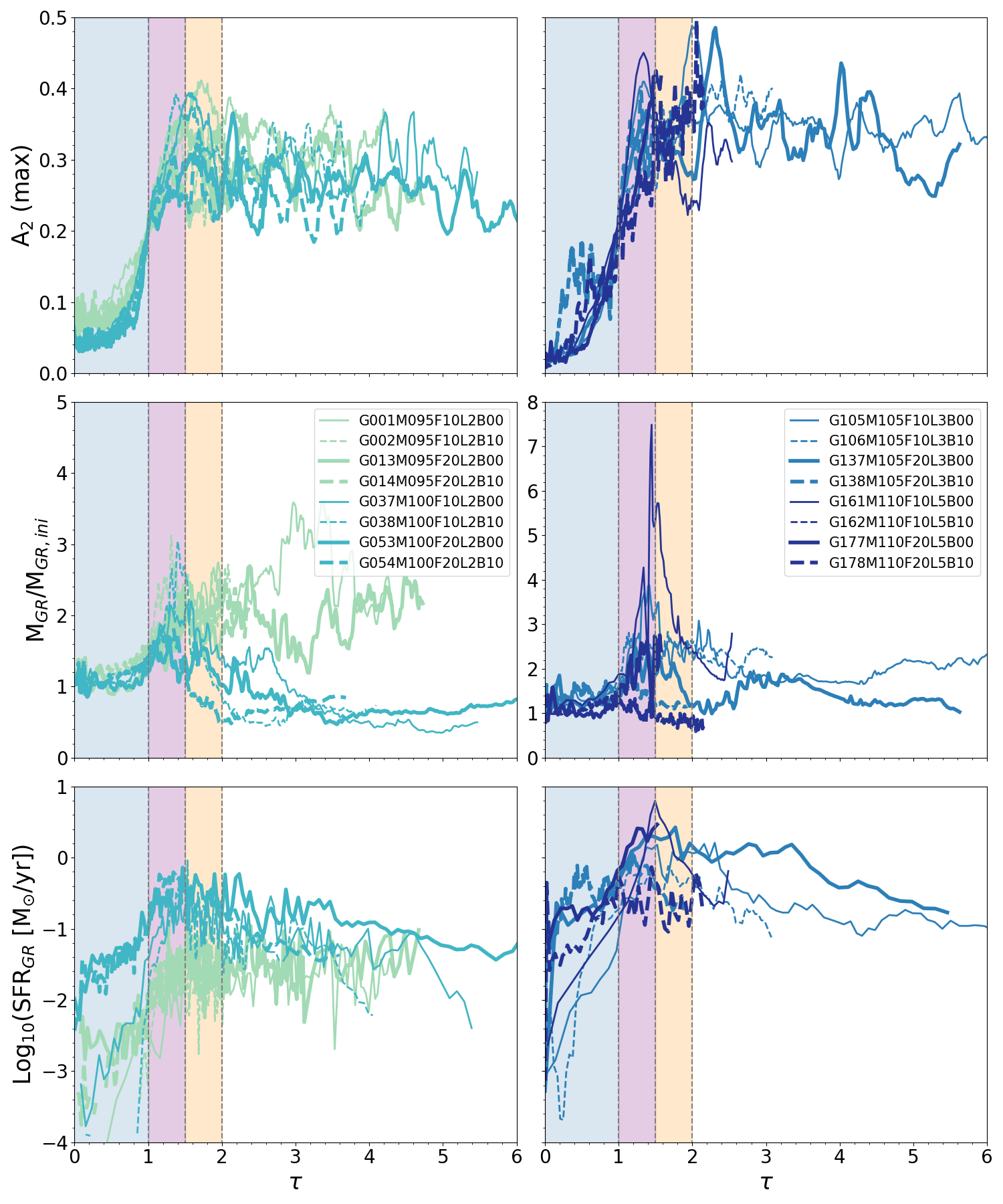}
\caption{Evolution of the maximum of the \A\ coefficient (top panels), the mass of gas within the central 1 kpc normalised by the initial mass of gas within the same radius (middle panels), and the SFR (bottom panels). The evolution is shown through the dimensionless parameter $\tau$, which is the ratio between the time of the simulations and the corresponding time when A$_{2}$ reaches the value of 0.2. The dashed vertical lines and shaded coloured areas show peculiar values of $\tau$ we use to describe the phases of the fuelling (i.e, $\tau \in$ 0-1 (blue area); $\tau \in$ 1-1.5 (purple area); $\tau \in$ 1.5-2 (orange area)). The colour code and the meaning of the different lines are the same as in Fig. \ref{fig:GR_BAR_SFR}.}
\label{fig:GR_BAR_SFR_norm}
\end{figure*}

Since the characteristic bar formation time seems to play a major role in the formation of the central gas reservoir, we have normalised the time by \tb\ and have introduced the dimensionless parameter $\tau = t/$\tb. We have also normalised the gas mass by the initial amount of gas contained within 1~kpc as shown in Fig.~\ref{fig:GR_BAR_SFR_norm}. In this figure, we can see that the bars and SFRs share a similar time evolution with a slight trend of increasing amplitude as a function of the stellar mass, for all simulations with initial stellar masses equal or above $10^{10}$~\Msun \ (see Sect.~\ref{subsec:regimes} for a discussion on simulations in the lower stellar mass bin). These similarities between barred systems confirm the relevance of the choice of $\tau$ allowing us to naturally account for the delay caused by the presence of a bulge. 

Except for the models G162M110F10L5B10 and G178M110F20L5B10, the normalised gas mass always peaks with $\tau$ between 1 and 1.5. This peak also roughly coincides with the maximum of the bar strengths and the SFRs. Those are associated with a central starburst leading to an increase in the number of new stars and a decrease in the gas mass inside the central 1 kpc. For simulations with M$_{\star} \geq 10^{10}$~M$_{\odot}$, we can roughly decompose the time evolution of the gas mass inside 1 kpc in three main phases referenced by the bar formation timescale \tb:
\begin{enumerate}
    \item From $\tau = 0$ to $\sim 1$, the bar forms and the gas mass inside the central 1~kpc region is nearly constant or increase weakly. The star formation rate slowly increases with time (by about 1 order of magnitude in 500~Myr).
    \item From $\tau = 1$ to $\sim 1.5$, the bar is strong (by definition, $A_2 > 0.2$) and significantly influences the redistribution of gas. Gas is being funneled towards the centre, leading to a more rapid increase of the gas mass, that triggers a starburst inside that region. The star formation evolution then reaches a plateau.
    \item From $\tau = 1.5$ to $\sim$2, the bar keeps its high amplitude, and a steeper decrease in the amount of gas in the central 1~kpc region is associated with a slow decline in star formation rate. This gas depletion phase marks the emergence of a small gaseous and stellar central mass overdensity, a discy structure that is fuelled by the bar. Follow-up star formation tends to predominantly occur within this inner disc region building up further beyond $\tau = 2$.
\end{enumerate}
The fact that those phases are similar for all barred models having an initial stellar mass bigger or equal to 10$^{10}$ \Msun suggests that the fuelling and the consumption of gas inside the central 1~kpc region is driven by the same physical phenomena and evolution, and only depends weakly on the initial gas fraction and the presence or absence of a central ellipsoid (within the range of parameters probed by our simulations). While there are clear local differences between simulations, it is still an interesting and relevant result as it sets the stage of the evolution of barred systems which could serve as an "isolated case" reference for further studies. Note that we are not here discussing the evolution beyond 3~Gyr, which could, for instance, lead to a secondary growth of the stellar bar, and to the emergence of large bars \citep[see e.g.][]{Schinnerer2023, Bland2023}.

\subsection{Gas structure in the central 1~kpc} \label{subsec:inner}
\begin{figure*}[h!]
\centering
\includegraphics[width=17cm]{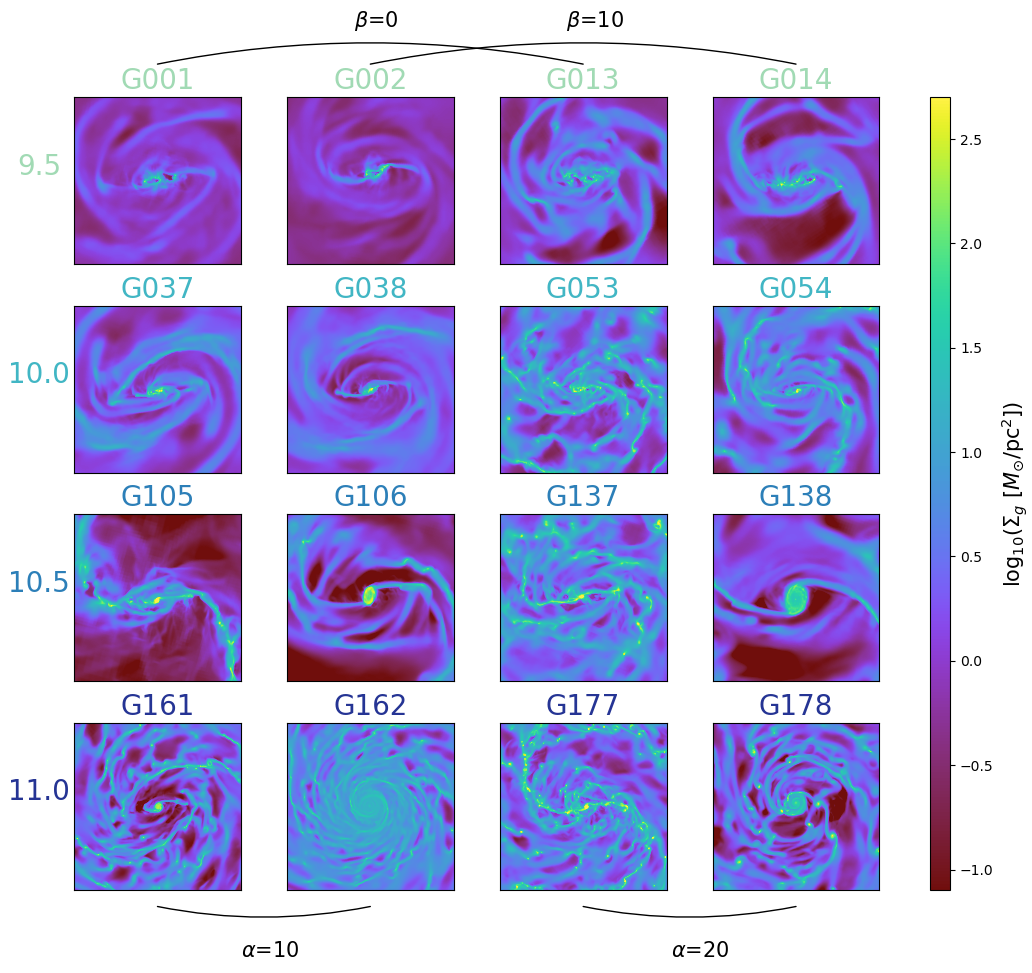}
\caption{Density map of gas for the 16 simulations at the corresponding time for which $\tau$ = 2. The first three rows show a box of 10 kpc side while the last row shows a box of 20 kpc side. The model numbers are colour-encoded according to the corresponding stellar mass (from the lightest to the most massive from the top to the bottom, i.e. 9.5, 10, 10.5, and 11 in log$_{10}$(\Msun)). The two left columns illustrate the model with a gas fraction of 10\% while the two right columns illustrate the models with a gas fraction of 20\%. The odd (even) reference number accounts for models without (with) a bulge.}
\label{fig:Sig_16_t2}
\end{figure*}

In Sect.~\ref{subsec:central_kpc}, we briefly discussed the property and global evolution in the central kpc. In Fig.~\ref{fig:Sig_16_t2}, we now present the gas distributions and reservoir by providing a gas density map for all 16 simulations at $\tau = 2$, in order to illustrate the emergence of distinct inner gas structures. In that figure, the four lines correspond to the four different stellar mass bins (colour-coded labels), the models having a lower (resp. higher) gas fraction ($\alpha$) being in the two left (resp. right) columns. The models without bulges ($\beta = 0$) are given by the first and third columns, whereas the models with bulges (here $\beta = 10$) are given by the second and the fourth columns. 

We observe a distinct central concentration of gas for almost all barred models having a stellar mass larger than 10$^{9.5}$ M$_{\odot}$. For these stellar mass ranges, only model G162M110F10L5B10 does not show any bar structure and central gas concentration. We see that models with a bulge have a more extended gas reservoir. While the bulge delays the bar formation, once the bar is formed, the bulge does not prevent the formation of a gas reservoir and actually allows the emergence of a prominent inner structure. We discuss the detailed properties and growth of those structures in a subsequent paper, but we can already suggest here that the extent of such an inner disc (or rings, see Fig.~\ref{fig:Sig_16_t2}) closely follows the change of the inner mass concentration.

\subsection{The conditional onset of inner stellar discs} \label{subsec:regimes}

As illustrated in Fig.~\ref{fig:GR_BAR_SFR}, we do not witness strong differences between simulations above and below the $10^{10}$~\Msun\ values in terms of the amplitude of the bar ($A_2$) or the global SFR. The main difference between those two sub-samples shows up in the evolution of the gas in the central 1 kpc. The decrease of the central gas mass leading to a long-term gas depletion mentioned as Phase~3 in the previous Section is observed for all galaxies with an initial mass of $10^{10}$~M$_{\odot}$ and above. For the less massive galaxies (i.e. models G001M095F10L2B00 , G002M095F10L2B10, G013M095F20L2B00 and G014M095F20L2B10, having an initial stellar mass of 10$^{9.5}$ M$_{\odot}$), the gas mass stays roughly constant or even increases during that phase. This means that the channelling of gas towards the central region (and subsequent star formation) is still occurring but seems to proceed in a different manner. 

To understand this better, we turn to the spatial distribution of star formation (and new stars). Fig.~\ref{fig:Dens_Glob_095} in the Appendix shows (second row in each panel) that for the lowest stellar mass models, star formation occurs mainly along and inside the bar until the end of the simulation without showing the emergence of a central gas reservoir. For the higher stellar mass models (Fig.~\ref{fig:Dens_Glob_100}, \ref{fig:Dens_Glob_105} and \ref{fig:Dens_Glob_110}), star formation occurs also along and inside the bar until $\tau=1.5-2$, when we start to see the emergence of a central gas overdensity. Subsequent star formation is mainly distributed inside this central gas reservoir. This result also connects with the fact that distinct inner streaming lanes, as well as disc or ring structures form and grow for the more massive end systems, while the lower mass systems exhibit more heterogeneous gas distribution within the bar region. This is quite an intriguing result as it means that star formation is either prevented or triggered in significantly different regions for the lowest mass simulation bin.

This trend has been reported by \citet{FraserMcKelvie2020} using emission-line mapping via integral-field spectroscopic observations for a large sample of nearby galaxies \citep[but see also][]{DiazGarcia2020}. The stellar mass value of $10^{10}$~\Msun\ at which this occurs also coincides with the one in our simulation set. It is worth mentioning that most of the galaxies presented in the above survey have a much higher gas fraction compared with the PHANGS sample. Their typical gas fraction \footnote{Value added Catalogs (SDSS): 
https://data.sdss.org/sas/dr17/-env/MANGA\_HI} in the 10$^{9.5-10}$ \Msun \ stellar mass bin is above $\sim$ 30~\% and goes up to $\sim$ 60-70~\%. Since we tuned our initial conditions to an observed sample of star-formation main sequence galaxies, this may mean either that the change is built in our set of morphological parameters (e.g. scale lengths) or that it is related to the varying relative contributions of physical processes at play, for example, the influence of feedback versus the strength of the gravitational potential, associated with a change of stellar mass. 

The physical origin of this change in the regime between the lowest and highest stellar masses is not yet fully understood. \citet{FraserMcKelvie2020} extrapolated from earlier simulations \citep{Emsellem2015} that shear may be an important ingredient in setting up such differences.
While this may play a role, we note that two of our simulations, one at $10^{9.5}$~\Msun\ and one at $10^{10}$~\Msun, share almost exactly the same radial mass gradient (only the mass scaling is different), and those two do present the above-mentioned change in the evolved morphology. We thus suggest that the main driver for such a difference lies with the relative contribution of the stellar-driven feedback within those scaled gravitational potential \citep[see e.g.][and references therein]{CollinsRead2022}. This will be specifically discussed in a subsequent paper.



\section{Summary and conclusion} \label{sec:sumandconc}
In this work, we have performed a set of 16 three-dimensional high-resolution hydro-dynamical simulations (see Table~\ref{Tab:Grid_16}) using the RAMSES AMR code to study and characterise the building and evolution of central gas reservoir in nearby main sequence disc galaxies. We have designed this grid based on the PHANGS-ALMA sample. We made use of four control parameters (i.e. stellar mass, gas fraction, scale length for the star distribution, and bulge mass fraction), and for each model started from axisymmetric initial conditions including gas, stars and dark matter.

We have further quantified the characteristic formation time for the bar \tb\ in our simulations using a reference value of 0.2 for \A, and compared them with the relation suggested by \citet{Fujii2019}. We have found that our simulations are located significantly below the relation (have smaller \tb \ at fixed $f_{disc}$). We could not conclude robustly on the origin of such a discrepancy as it may both reflect the intrinsic scatter of such a relation and the imposed variety in the initial conditions (i.e. halo concentration, anisotropy). A larger set of simulations including gas and star formation covering lower values of f$_{disc}$ (hence larger values of \tb) are needed to confirm this trend.

We have found that models G162M110F10L5B10 and G178M110F20L5B10 are expected to be significantly more stable against bar formation (according to the criterion established by \citet{Efstathiou1982}; but see Sect.~\ref{sec:formtime}). Note that the criterion we use to decide if a bar is formed or not (i.e. \A \ = 0.2) is met for model G178M110F20L5B10 despite the lack of an apparent proper bar structure: we interpret that case as \A\ capturing the evolution of the strong spiral arms.

We have studied the impact of three control parameters on the evolution of the central 1~kpc region. The mass inside the central gas reservoir naturally increases with the initial stellar and gas mass. The presence of a bulge delays the formation of the bar (i.e. values of \tb \ are larger) and thus the formation of the gas reservoir, but does not prevent its formation.

The global evolution of the 12 models having a stellar mass $\geqslant$~10$^{10}$ M$_{\odot}$ can be roughly described using a dimensionless bar formation time parameter $\tau=t$/\tb, including three subsequent phases:
\begin{enumerate}
    \item A formation phase: from $\tau = 0$ to $\sim 1$, the bar forms and the gas mass inside the central 1~kpc region is nearly constant or increases weakly. The star formation rate slowly increases with time.
    \item A fueling and growth phase: from $\tau = 1$ to $\sim 1.5$, the bar is strong enough (i.e. $A_2 > 0.2$) and starts to transport gas towards the centre, leading to a steeper increase of the gas mass and a starburst inside that region. The star formation reaches a plateau.
    \item A depletion phase: from $\tau = 1.5$ to $\sim$2, the bar stays strong, and a steep decrease in the amount of gas in the central 1~kpc region is associated with a slow decline in star formation rate. This phase witnesses the emergence of a central stellar mass seed growing into a more extended inner stellar and gas structure (discs and rings). The sizes of the inner discy structure seem to vary with e.g. the initial stellar mass and the presence or absence of an ellipsoid.
\end{enumerate}

Simulated galaxies with initial stellar masses below $10^{10}$~\Msun\, falling in the lowest mass bin, exhibit differences with respect to the more massive galaxies in the sample as for their inner gas structures, the spatial distribution of star forming regions for $\tau \ge 2$ and the gas depletion timescales in the central kpc. More specifically, the two first above-mentioned phases are also witnessed for the lower stellar mass models (M$_{\star}$=10$^{9.5}$ M$_{\odot}$), but we do not observe Phase~3, that is, a steep decrease of the gas mass inside the 1~kpc central region at $\tau > 1.5$. The gas mass in the central kpc thus stays roughly constant and is associated with a relatively constant SFR over time (until at least $\tau$=4). We have also shown evidence for two distinct star formation distributions for $\tau \ge 2$ below and above the $10^{10}$~\Msun\ initial stellar mass of our models. For models with M$_{\star} < 10^{10}$ M$_{\odot}$ (hence only for the lower mass bin),  we observe that the star formation mainly occurs along the bar, whereas star formation is mostly triggered inside the inner gas concentration for models with M$_{\star}$ $\geq$ 10$^{10}$ M$_{\odot}$. This trend echoes the reporting made by \citet{FraserMcKelvie2020} using H$\alpha$ two-dimensional mapping of a sample of nearby galaxies. We suggest that this relates to the relative contribution of stellar-driven feedback within a given gravitational potential: this will be probed and discussed in detail in a subsequent paper (Verwilghen in preparation). The subset of simulations presented here only covers a restricted range of observational properties and are constrained by a fixed and limited set of initial structural parameters. We can thus already presume that stellar mass may not be the only or even the prime driver of the differences we observe in our simulations \citep[see e.g.][]{DiazGarcia2020}.

In this paper, we provide a first pilot study at the structures and time evolution of a set of simulations probing the star formation main sequence of nearby disc galaxies. Subsequent papers in this series will focus on examining in detail the advent and growth of the above-mentioned inner gas structures (and their associated stellar content), as well as study in more detail the origin of the change of regime around $10^{10}$~\Msun\ observed in \citet{FraserMcKelvie2020} and reproduced with this first subset of models. We have already planned for an extended sample of 54 models better covering the range of observed properties of the PHANGS-ALMA sample in this stellar mass range: a more full-fledged account of this "complete" set of hydro-dynamical simulations will be presented when available. Such studies can then serve as a benchmark for simulations embedded in a more comprehensive environment, including, for example, gas accretion, interactions or evolution in a cosmological context.
%

\begin{acknowledgements}
We thank the anonymous referee for a report that helped clarify this manuscript. PV acknowledges support from the Excellence Cluster ORIGINS which is funded by the Deutsche Forschungsgemeinschaft (German Research Foundation) under Germany’s Excellence Strategy – EXC-2094–390783311. The simulations in this paper have been carried out on the computing facilities of the Computational Center for Particle and Astrophysics (C2PAP). We are grateful for the support by Alexey Krukau and Margarita Petkova through C2PAP. 

FR acknowledges support provided by the University of Strasbourg Institute for Advanced Study (USIAS), within the French national programme Investment for the Future (Excellence Initiative) IdEx-Unistra. 

MV is supported by the Fondazione ICSC National Recovery and Resilience Plan (PNRR), Project ID CN-00000013 `Italian Research Center on High-Performance Computing, Big Data and Quantum Computing' funded by MUR - Next Generation EU. MV also acknowledges partial financial support from the INFN Indark Grant.

RSK acknowledges financial support from the European Research Council via the ERC Synergy Grant `ECOGAL' (project ID 855130),  from the German Excellence Strategy via the Heidelberg Cluster of Excellence (EXC 2181 - 390900948) `STRUCTURES', and from the German Ministry for Economic Affairs and Climate Action in project `MAINN' (funding ID 50OO2206). RSK thanks for computing resources provided by the Ministry of Science, Research and the Arts (MWK) of the State of Baden-W\"{u}rttemberg through bwHPC and the German Science Foundation (DFG) through grants INST 35/1134-1 FUGG and 35/1597-1 FUGG, and also for data storage at SDS@hd funded through grants INST 35/1314-1 FUGG and INST 35/1503-1 FUGG.

KD acknowledges financial support from `BiD4BEST' - European Innovative Training Network (ITN) funded by the Marie Sk\l{}odowska-Curie Actions (860744) in Horizon 2020 and by the COMPLEX project from the European Research Council (ERC) under the European Union’s Horizon 2020 research and innovation program grant agreement ERC-2019-AdG 882679.

JN acknowledges funding from the European Research Council (ERC) under the European Union’s Horizon 2020 research and innovation programme (grant agreement No. 694343).

KG is supported by the Australian Research Council through the Discovery Early Career Researcher Award (DECRA) Fellowship (project number DE220100766) funded by the Australian Government. KG is supported by the Australian Research Council Centre of Excellence for All Sky Astrophysics in 3 Dimensions (ASTRO~3D), through project number CE170100013. 

MCS acknowledges financial support from the European Research Council under the ERC Starting Grant `GalFlow' (grant 101116226) and from the Royal Society (URF\textbackslash R1\textbackslash 221118).

\end{acknowledgements}

\bibliographystyle{aa} 
\bibliography{main_aa} 
%
%

\begin{appendix} 
\section{Estimation of the fit values}
\begin{figure}[b!] \label{fig:fits}
\centering
\includegraphics[width=8.cm]{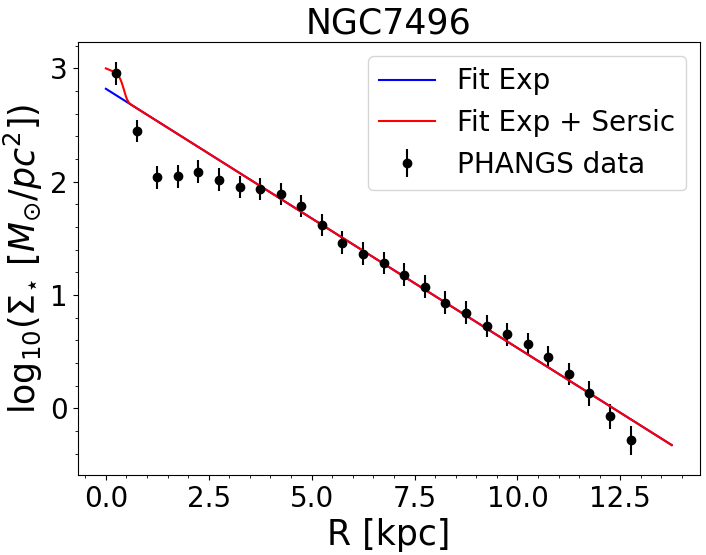}
\hfil
\includegraphics[width=8.2cm]{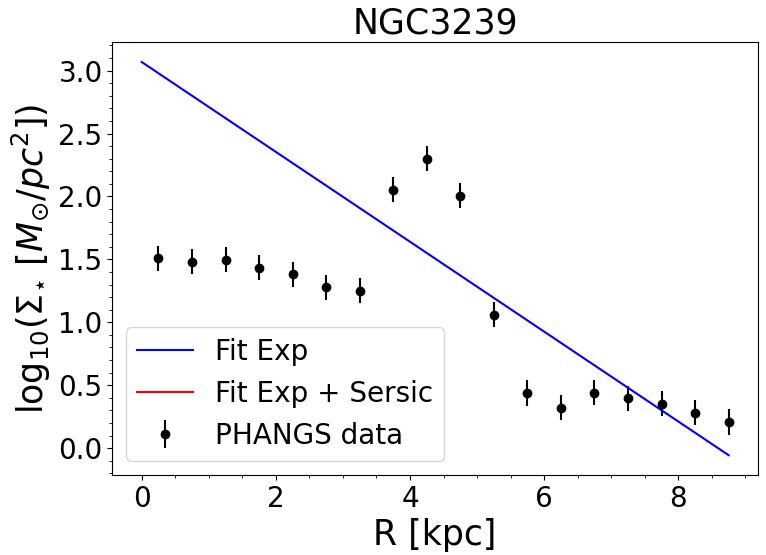}
\caption{Example of fits for which we attributed the value 0 (satisfying but with significant residuals; top panel) and -1 (bad; bottom panel).}
\label{fig:Fit_0_1}
\end{figure}

As introduced in Sect.~\ref{sec:meth} and illustrated in Fig.~\ref{fig:fit}, we used a subjective rating process to sort the different results coming from the exponential and Sersic fits of the 1D stellar and gas density profiles. Fig.~\ref{fig:Fit_0_1} briefly illustrates this rating for two profiles that were attributed the values 0 and -1, and thus considered as `satisfying' and `bad', respectively. The left panel (NGC\,7476) represents the case for which we considered the fit to be satisfying (value 0) and the right panel (NGC\,3239) represents the case we considered the fit to be bad (value -1). The `bad' case shows one example of a density profile for which the adopted functional form (exponential plus Sersic) is clearly not appropriate. While this classification is subjective, we were mostly interested in the global trends, and it did not strongly influence the choice of the control parameter values themselves (see Fig.~\ref{fig:PH_IC_Tot}). 
\section{Evolution of the models as a function of the parameter $\tau$} \label{app:Evol_tau} 
\begin{figure*}[h!]
\centering
\includegraphics[width=17cm]{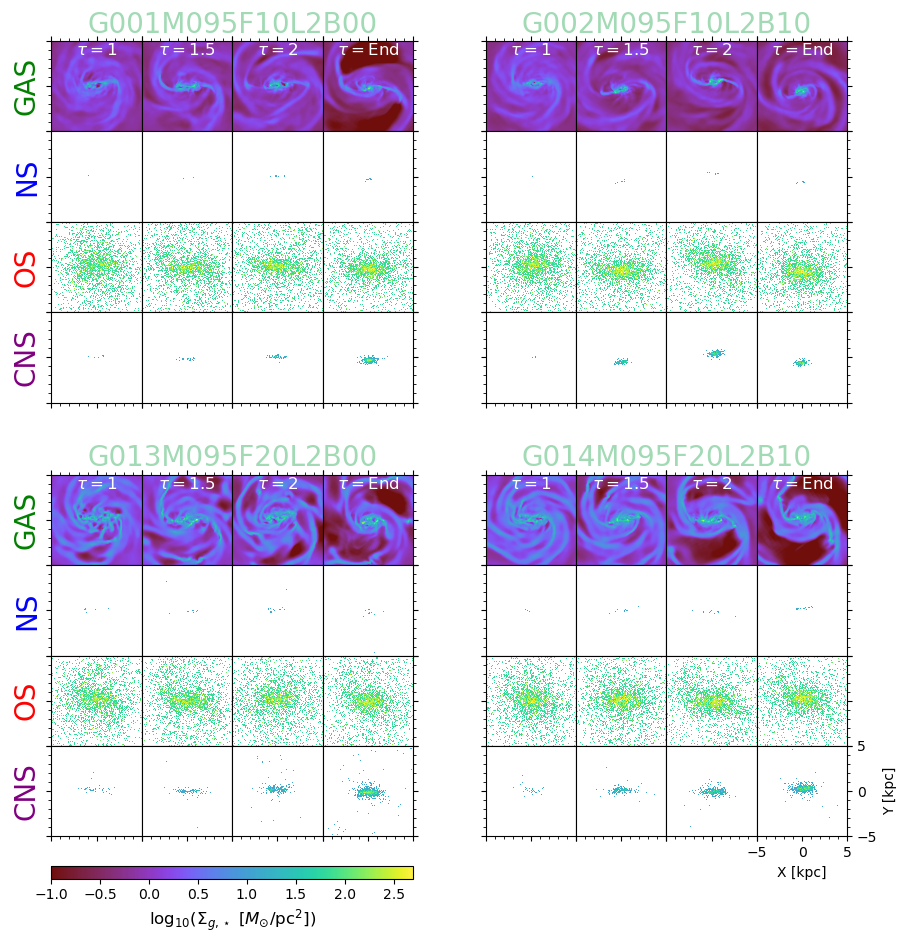}
\caption{Time evolution as a function of $\tau$ of the gas surface density (top row, \textcolor{OliveGreen}{GAS}), formed stars' surface density (second row, \textcolor{blue}{NS}), old stars' surface density (third row, \textcolor{red}{OS}), and cumulative formed stars' surface density (bottom row, \textcolor{violet}{CNS}) of the four lowest stellar mass models (10$^{9.5}$ \Msun). Each panel shows a box with a side length of 10 kpc.}
\label{fig:Dens_Glob_095}
\end{figure*}

\begin{figure*}[h!]
\centering
\includegraphics[width=17cm]{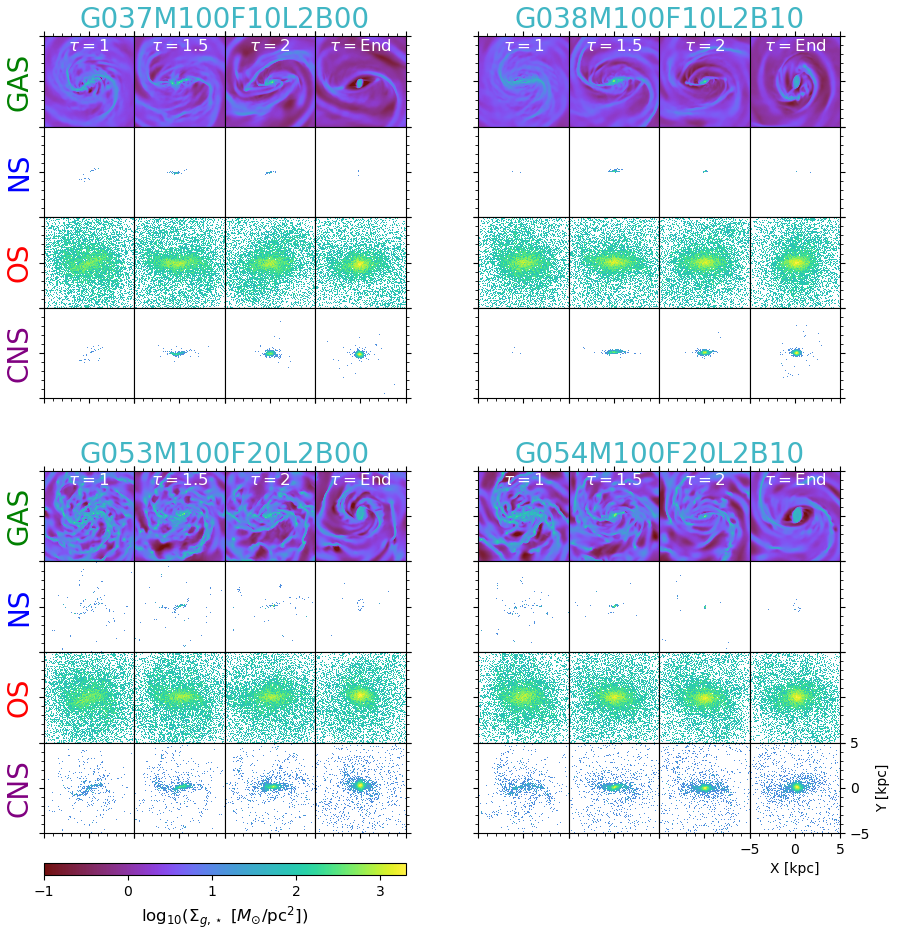}
\caption{Same as Fig.~\ref{fig:Dens_Glob_095} for the four 10$^{10}$ \Msun\ stellar mass models.}
\label{fig:Dens_Glob_100}
\end{figure*}

\begin{figure*}[h!]
\centering
\includegraphics[width=17cm]{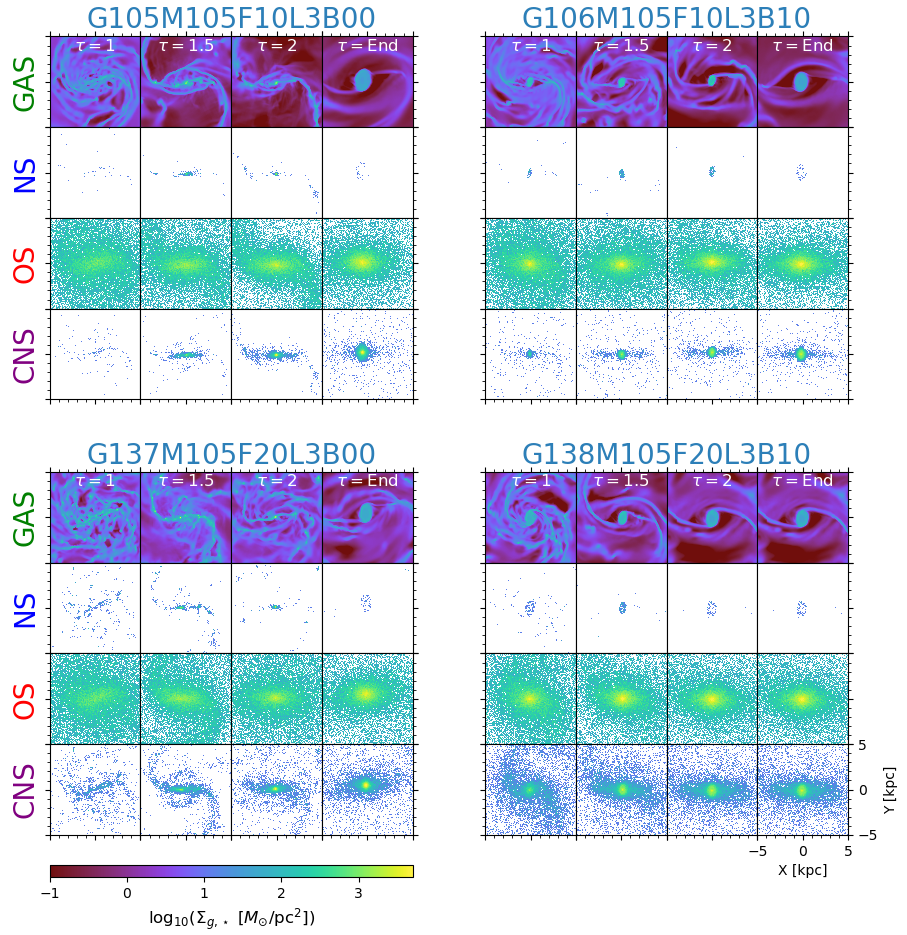}
\caption{Same as Fig.~\ref{fig:Dens_Glob_095} for the four 10$^{10.5}$ \Msun\ stellar mass models.}
\label{fig:Dens_Glob_105}
\end{figure*}

\begin{figure*}[h!]
\centering
\includegraphics[width=17cm]{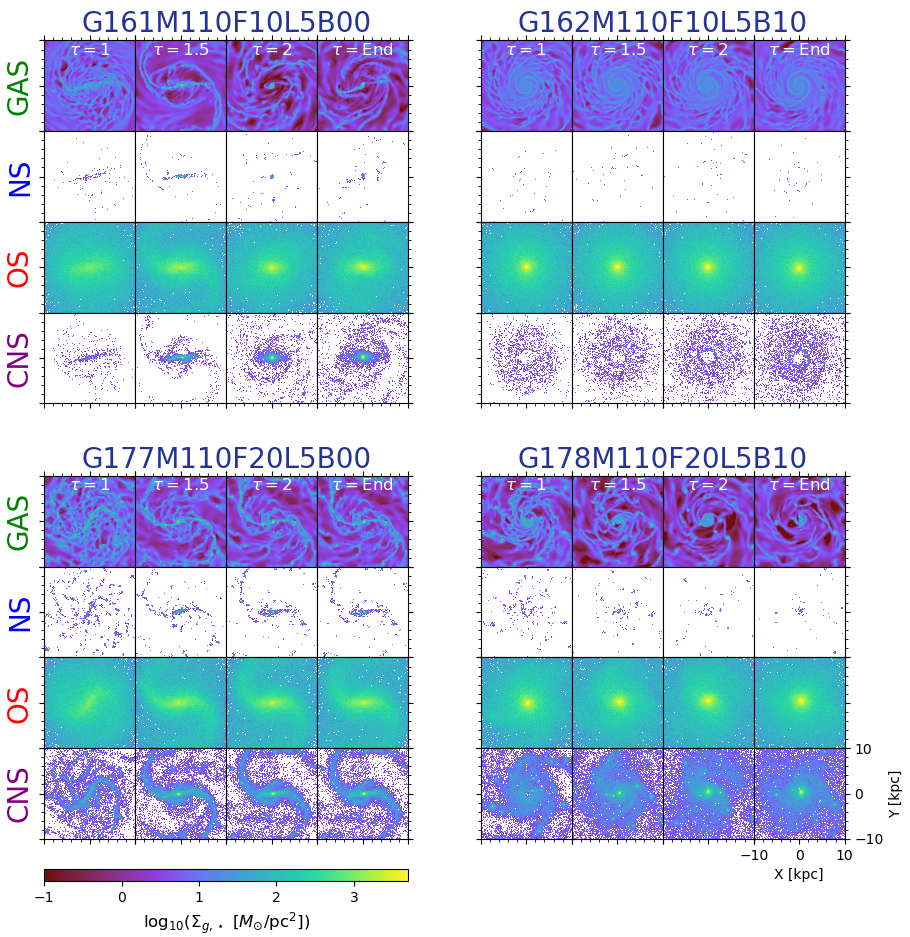}
\caption{Same as Fig.~\ref{fig:Dens_Glob_095} for the four 10$^{11}$ \Msun\ stellar mass models. The box size has been extended to 20~kpc for this most massive stellar bin.}
\label{fig:Dens_Glob_110}
\end{figure*}

As mentioned in Sect.~\ref{subsec:phases}, we have identified three distinct phases in the building and evolution of the gas reservoir by using the dimensionless parameter $\tau$ = t/\tb, where \tb \ is the typical bar formation time. The following figures illustrate the time evolution of the system for the 16 simulations providing snapshots at four different values of $\tau$ (i.e. 1, 1.5, 2, and End, for the end of the simulation). For each simulation, we gather the four models belonging to the same initial stellar mass into one figure. Each panel corresponds to one labelled simulation: each include a zoom (box of 10 kpc on a side) and 4 row with the surface densities of (from top to bottom) the gas, of the (new) stars formed within 50 Myr of the corresponding snapshot time, of all old (initial) stars and of all (new) stars (formed since the beginning of the simulation). Fig.~\ref{fig:Dens_Glob_095}, \ref{fig:Dens_Glob_100}, \ref{fig:Dens_Glob_105}, \ref{fig:Dens_Glob_110} refer to the models with a stellar mass of 9.5, 10, 10.5 and 11 log$_{10}$(\Msun), respectively. 

In the top rows of the most massive initial stellar mass bins (Fig.\ref{fig:Dens_Glob_100}, \ref{fig:Dens_Glob_105}, \ref{fig:Dens_Glob_110}) we witness the formation of a clearly marked inner gas structure (except for model G162M110F10L5B10, and for model G178M110F20L5B10), while simulations presented in Fig.~\ref{fig:Dens_Glob_095} do not exhibit such gas structures. We also observe a change in the distribution of new stars that tend to emerge along the bar structure for the lowest stellar mass models, and are more localised (inner region) for the higher stellar mass models from $\tau \ge 2$. There is an associated difference in the structure of the bar with the more massive galaxies presenting a more distinct and rounder stellar concentration (within the bars). This central concentration could be caused by the emergence of an ILR as discussed in Sect.~\ref{subsec:building_bar}, and ~\ref{subsec:regimes}.
\end{appendix}
\end{document}